\documentclass[9pt,twocolumn,twoside]{osajnl}

\journal{josaa} 
\usepackage{comment}

\usepackage{graphicx}
\usepackage{caption}
\usepackage{subcaption}
\usepackage{setspace}

\DeclareCaptionLabelSeparator{emdash}{\textemdash}
\setboolean{shortarticle}{false} 

\ifthenelse{\boolean{shortarticle}}{\colorlet{color2}{color2b}}{\colorlet{color2}{color2}} 

\title{High-contrast self-imaging with ordered optical elements}

\author[1,*]{Ali Naqavi}
\author[2]{Hans Peter Herzig}
\author[3]{Markus Rossi}

\affil[1]{Thomas J. Watson Laboratories of Applied Physics, California Institute of Technology, Pasadena, California 91125, USA}
\affil[2]{Optics \& Photonics Technology Laboratory, 
\'{E}cole Polytechnique F\'{e}d\'{e}rale de Lausanne, 
Rue de la Maladi\`{e}re 71b, CH-2000 Neuch\^{a}tel, Switzerland}
\affil[3]{Heptagon, Moosstrasse 2, CH-8803 R\"{u}schlikon, Switzerland}

\affil[*]{Corresponding author: naqavi@caltech.edu}

\dates{Compiled \today}

\ociscodes{(110.0110) Imaging systems; (070.0070 ) Fourier optics and signal processing;(050.0050) Diffraction and gratings.}

\doi{\url{http://dx.doi.org/10.1364/ao.XX.XXXXXX}}

\begin{abstract}
Creating arbitrary light patterns finds applications in various domains including lithography, beam shaping, metrology, sensing and imaging. 
We study the formation of high-contrast light patterns that are obtained by transmission through an ordered optical element based on self-imaging.
By applying the phase-space method, we explain phenomena such as the Talbot and the angular Talbot effects. 
We show that the image contrast is maximum when the source is either a plane wave or a point source, and it has a minimum for a source with finite spatial extent. We compare these regimes and address some of their fundamental differences. 
Specifically, we prove that increasing the source divergence reduces the contrast for the plane wave illumination but increases it for the point source. 
Also, we show that to achieve high contrast with a point source, tuning the source size and its distance to the element is crucial.
We furthermore indicate and explore the possibility of realizing highly complex light patterns by using a periodic transmission element. These patterns can have more spots in the far field than the number of diffraction orders of the periodic element. 
We predict that the ultimate image contrast is smaller for a point source compared to a plane wave.
Our simulations confirm that the smallest achievable spot size in the image is imposed by diffraction regardless of the imaging regime. 
Our research can be applied to similar domains e.g. quantum systems.
\end{abstract}

\setboolean{displaycopyright}{true}

\begin{document}

\maketitle
\thispagestyle{fancy}

\ifthenelse{\boolean{shortarticle}}{\ifthenelse{\boolean{singlecolumn}}{\abscontentformatted}{\abscontent}}{}

\section{Introduction}
\label{sec:1}

\begin{figure}
\begin{tabular}{c}
\includegraphics[width=0.25\textwidth]{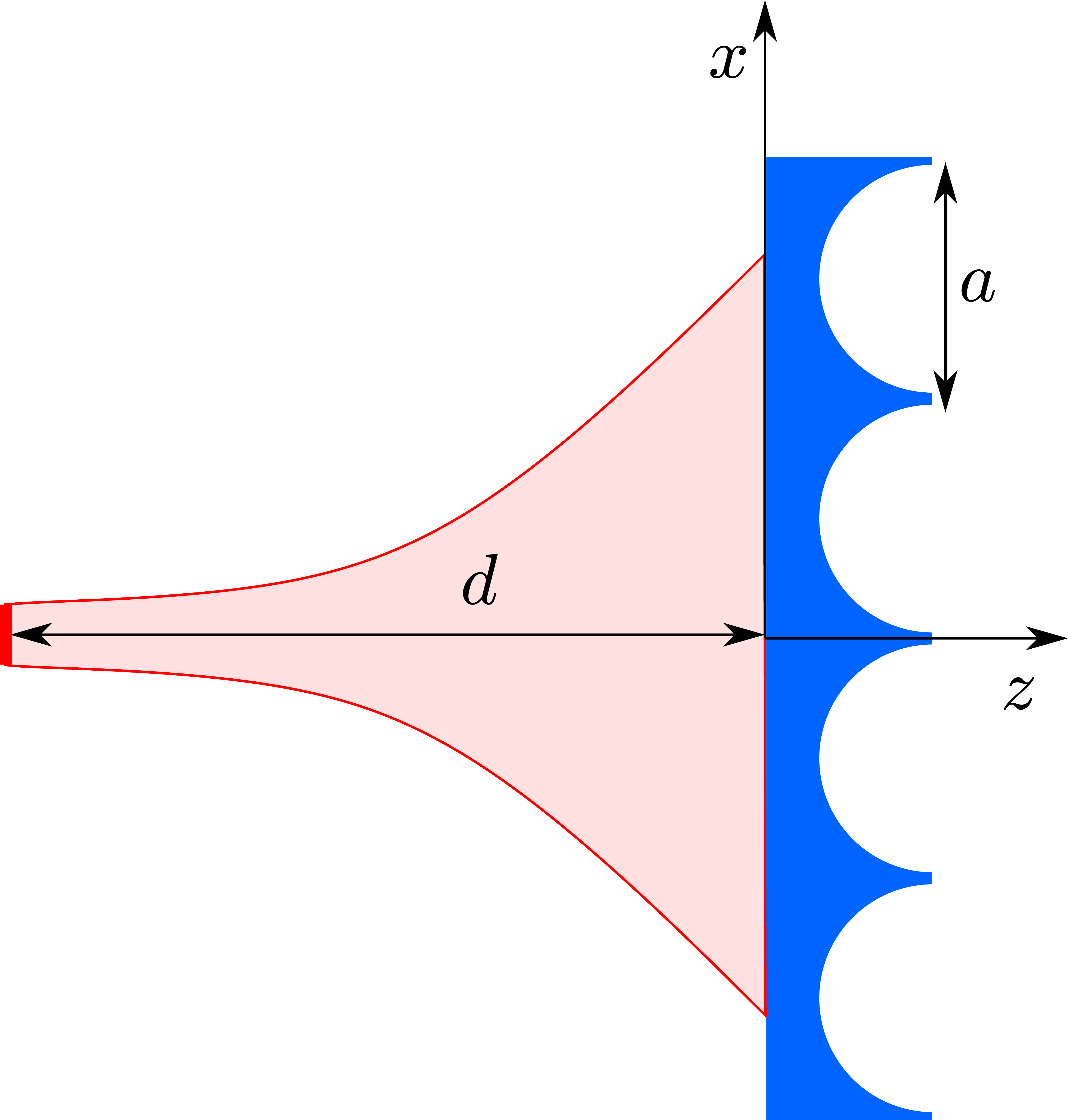}\\
\end{tabular}
\captionsetup{singlelinecheck=off}\caption{The setup under study: a monochromatic light source with finite spatial extent illuminates an OTE. Often the image at $z\rightarrow\infty$ is of interest.}
\label{fig:0}
\end{figure}

Creating arbitrary high-contrast patterns by using optical waves is a requirement for the advancement of various technologies including microscopy, lithography, sensing, imaging and metrology. 
In a basic scenario that is depicted in Fig.\ref{fig:0}, an optical transmission element is illuminated by a monochromatic source to produce high-contrast images in the output. 
Several known effects can be appreciated to form an image with high contrast, e.g. Talbot and Montgomery, which are realized by using periodic and quasi-periodic elements respectively \cite{Berry, LohmannFractional}. 
These phenomena are usually referred to as \textit{self-imaging} effects. 
One should however note that random structures can create high-contrast images too, as evidenced by the existence of speckles.  
The Talbot effect occurs in the near-field region of a periodic Ordered Transmission Element (OTE) when it is illuminated by a plane wave. 
The order in the OTE refers to the configuration in which its individual elements are put and in general, it can be periodic, quasi-periodic, etc.
Examples of periodic OTEs include optical gratings, lens arrays, etc. 
In this manuscript the near-field refers to the region close to the OTE  where its arrayed structure is most pronounced and the effect of its spatial extent is negligible. At larger distances, the effect of array extent becomes important. The latter case is referred to as the far field here. 
An analogy can be found between our definition of the near- or the far-field regions, and the Fresnel or the Fraunhofer regions in classical optics.   However these definitions should not be associated to the concept of surface waves and plasmonics, which apart from the common term \lq\lq near field" have nothing to do with our investigations in this paper. 
In the far-field of the considered periodic optical OTE that is illuminated by a plane wave, the incident wave is separated in the form of diffraction orders. 
Similar effects have been predicted for a quasi-periodic OTE \cite{Berry,LohmannFractional}. 
Later it has been shown that if the periodic OTE is illuminated by a point source instead of a plane wave, the far-field image remains periodic but as a function of angle \cite{Som, testorf2010phase,Azana}, which has been addressed as the angular Talbot effect recently \cite{Azana}. These two particular cases can be generalized by considering a source with finite spatial extent and an arbitrary OTE, which does not need to be necessarily periodic. In this article we explore the passage of monochromatic light that is emitted from a source with finite spatial extent through a one-dimensional (1D) periodic OTE. We prove the existence of two regimes for high-contrast imaging that are fundamentally different, and we investigate their optical behavior and the criteria for the formation of a high-contrast image in each one of them. 

To engineer the image optical properties, appropriate analytical tools are erquired. 
One of the major bottlenecks in the investigation and design of imaging systems is their sophisticated analytical treatment, which eventually necessitates performing numerical simulation.
Due to this complexity in calculations, a main body of the literature concerning our problem of interest either applies to a very limited number of special cases, or hardly provides enough intuition \cite{Berry,LohmannFractional,Som,Azana}.
In classical optics calculations, often the response of an optical system is obtained by assuming illumination that is fully localized either in $k-$space (plane-wave), or in real space (point-source). 
The \textit{phase space} provides a means to represent the response of the system versus both the position and the wave vector, thus allows achieving deeper intuition into the system optical behavior \cite{SchleichQO,Bastiaans1,Bastiaans2} and more advanced optical design \cite{testorf2006designing}.  
We base our analysis in this manuscript on the phase-space interpretation of optical fields. 
The Talbot effect, the angular Talbot effect, the Montgomery effect and the Lau effect can all be explained with a minimal amount of computational effort by using this approach\cite{ojeda1985quasi,testorf2010phase}.   
The phase space approach gives us insight on the reason behind the dissimilar optical behavior of the system in different regimes and allows us predict its optical properties when new conditions are encountered. 
We find that the image contrast is larger for a point source or a plane wave, compared to a source with finite spatial extent.
Our study uncovers non-conventional behavior for each of these two regimes. For example, we show that depending the source divergence, contrast may be destroyed such that diffraction orders are not observable even at infinity. 
We confirm the crucial role of the distance between the source and the OTE, and the source divergence, and we find criteria for high-contrast optical self-imaging in the two regimes. 
Despite simplicity, the analytical method that we use can even provide insight into characteristics such as image contrast. For instance we show that the ultimate theoretical image contrast is larger for a plane wave compared to a point source. 

Our findings in this paper are based on several assumptions. 
In a main body of this article, we assume for simplicity that the problem is invariant in $y$ direction and the OTE is 1D periodic in $x$ direction. Nevertheless, our conclusions can be extended easily to 2D-ordered structures. 
Other types of order are discussed shortly in the last part of the paper. 
While treating the system analytically, we consider different source types, but for numerical simulations we assume that the source is a Gaussian beam with its waist at $z=0$. 
Light propagation after the OTE can be rigorously solved by using scalar approximation because the propagation occurs in free space. Due to the similarity of the scalar wave equation and the Schr{\"o}dinger equation, the results of our study also apply to the realm of quantum mechanics \cite{SchleichQO}.  
Another assumption is that the source propagation direction and the OTE normal are the same. Oblique incidence of light on an ordered structure is an interesting matter to study, but is beyond the scope of this article. 

To address our questions of interest, we provide a graphical model which is based on an analytical treatment of the optical problem. 
We found very recently a model that is introduced a few years ago by Testorf \cite{testorf2010phase}, which is similar to our model, apart from slight details. Testorf's model mainly explains self-imaging for either a plane wave or a point source. We explore this model more deeply and extend it to explain the reasons behind the loss of contrast for a source with finite extent. This also allows us to conclude criteria for high-contrast self-imaging. 
The predictions of this model are in agreement with our numerical simulations, which are based on the angular spectrum method.  
Section \ref{sec:2} describes the analytic model, based on which we derive criteria for high-contrast optical self-imaging. 
In section \ref{sec:4} we show that the theory mentioned in section \ref{sec:2} needs to be further developed to allow understanding of  point-source imaging. We thus complete it in section \ref{sec:5}. Sections \ref{sec:4} and \ref{sec:5} contain key messages in this article. They include the proof of existence and explanation of the two different regimes for optical self-imaging. 
Section \ref{sec:6} includes an extension of the method to more general types of order in the OTE, conclusions and final remarks.

\section{Theory of self-imaging in phase space}
\label{sec:2}
The 1D Wigner distribution function (WDF) of a monochromatic scalar field $\psi$ is defined as 
\begin{equation}\label{eq:wdf}
W(x,k_{x})=\frac{1}{2\pi}\int{dx^{'}\psi^{*}(x+\frac{x^{'}}{2})e^{ik_{x}x^{'}}\psi(x-\frac{x^{'}}{2})}.
\end{equation}
where $k_x$ is the wave vector along $x$ axis. 
Historically, the WDF was defined for the quantum treatment of thermal equilibrium \cite{Wigner}, but soon it was proved  to be very useful to solve optics problems too \cite{Bastiaans1,Bastiaans2,alonso2011wigner}.  

The WDF has interesting properties that distinguish it as an appropriate candidate for wave behavior analysis. Integrating it along the $x$ axis gives the wave intensity distribution in $k$-space, which is sometimes called the power spectral density. A similar integral over $k_x$ results in the intensity as a function of position $x$. Integration of the WDF over the whole phase space, i.e. over both $x$ and $k_x$,  gives the wave total power. Besides, the evolution of the wave WDF during propagation can be taken into account by simple geometric transformations. Within the paraxial regime, the propagation along $z$ axis over a distance $d$ can be modeled by a shear of $k_x d/k_0$  along $x$ axis, where $k_0$ is free space wave vector. Passage through the OTE is equivalent to multiplication in $x$ and convolution along $k_x$, and the propagation to infinity is equivalent to a $90$-degree rotation \cite{Lohmann}. 

\begin{figure}
\begin{tabular}{cc}
\includegraphics[width=0.2\textwidth]{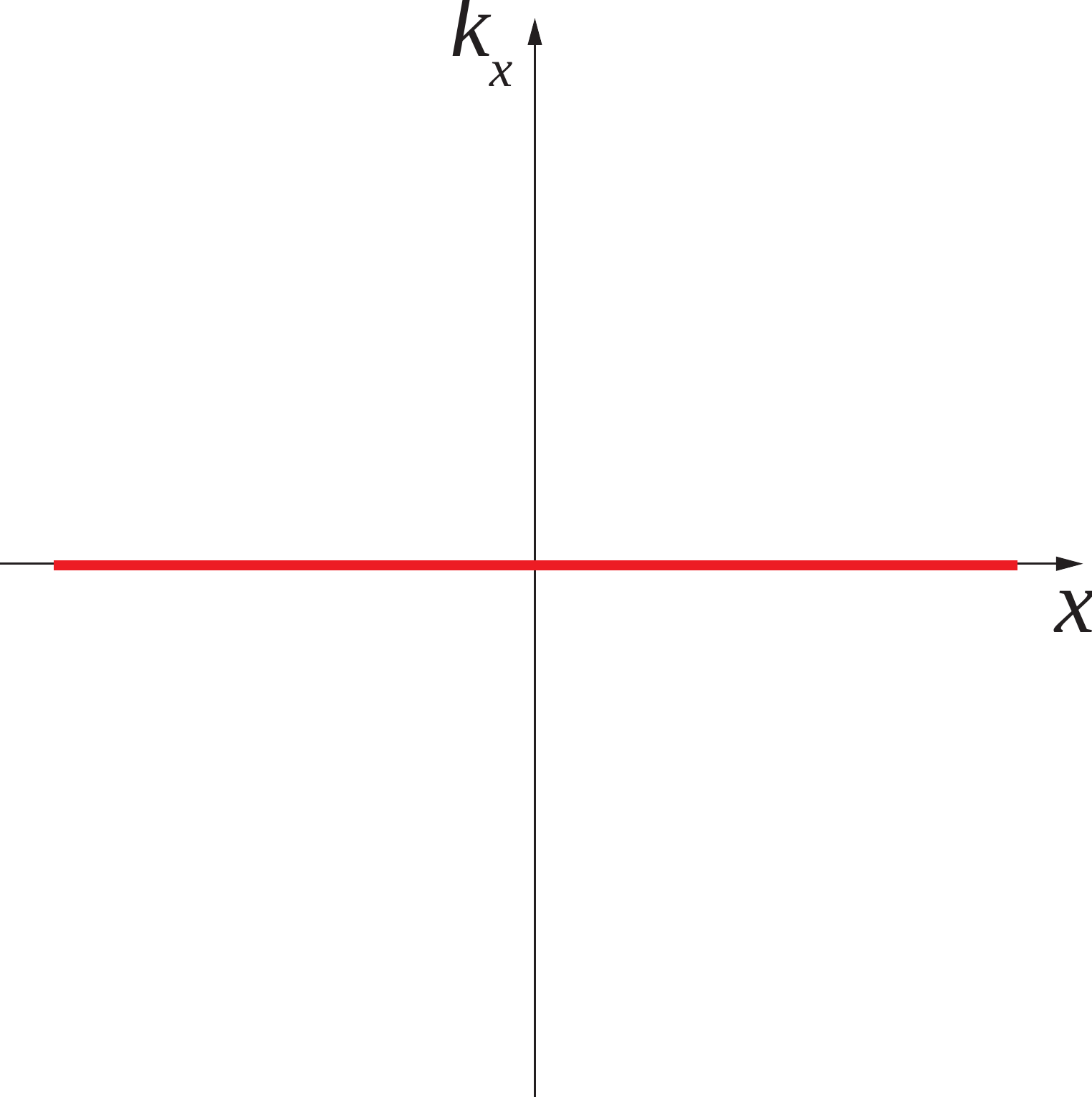} &
\includegraphics[width=0.2\textwidth]{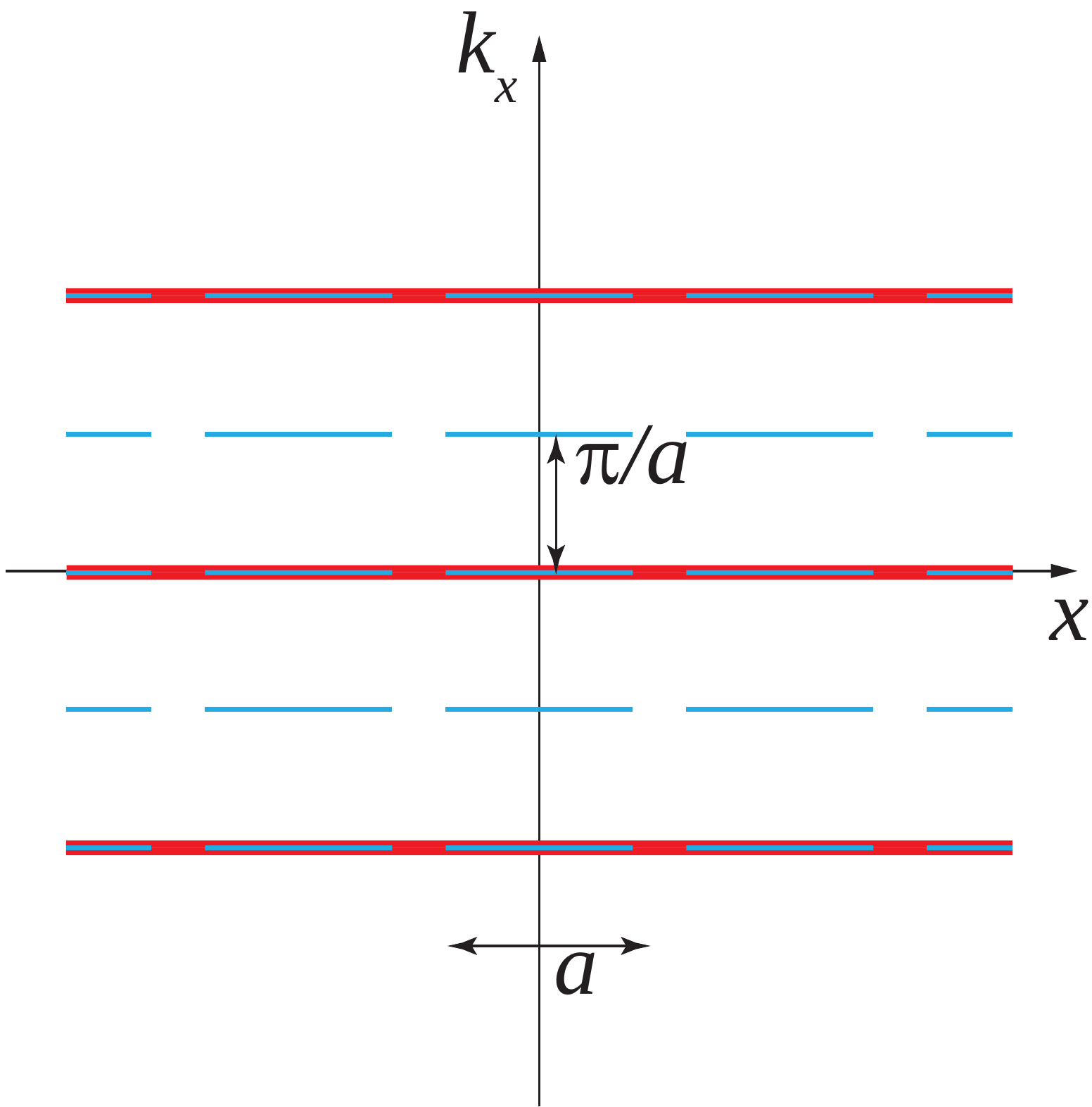}\\
(a) & (b) \\
\end{tabular}
\captionsetup{singlelinecheck=off}\caption{The WDF of (a): a plane wave that propagates along the optical axis ($z$). (b): a periodic OTE with period $a$.}
\label{fig:1}
\end{figure}

We consider different sources in our analysis: uniform plane wave, point source, general realistic source with the effective spread of $\Delta x$ and $\Delta k_x$ in real- and $k$-space, and a Gaussian source as a special example of the latter case. The first two sources are extreme cases where the wave’s power is completely concentrated at either a single spatial frequency ($k_x$) or a single position ($x$). Due to uncertainty, a realistic source always shows an intermediate behavior and has a certain extent over both real- and $k$-space. 
A plane wave pointing toward $z$ is represented by a line in phase space as indicated in Fig. \ref{fig:1} (a). 
Since in our case the OTE is periodic in $x$, its transfer function $t(x)$ can be expressed in terms of a Fourier series with Fourier coefficients $t_n$, and its WDF can be expressed as \cite{Schleich}
\begin{eqnarray}\label{eq:wdf0}
W(x,k_{x};0)=\frac{1}{2\pi}\sum_{mn}t_{n}^{*}t_{m}\exp\left\{ i\frac{2\pi x}{a}(m-n)\right\} \nonumber\\\times\delta(k_{x}-\frac{2\pi}{a}\frac{(m+n)}{2}).
\end{eqnarray}

Two types of terms appear in the WDF of Eq. \ref{eq:wdf0}; oscillatory with respect to $x$ ($m\neq n$) which are usually called \textit{intermodulation} in the literature and non-oscillatory ($m=n$). 
Fig.\ref{fig:1} (b)  shows these terms schematically as dashed and solid lines respectively. We mainly care about the largest period of the intermodulations terms which is $a$, because usually when it fulfills the requirements of high-contrast imaging, terms with smaller periods $a/(m-n)$ do so as well. 
All orders are plotted with the same intensity but for a real structure they should be weighted according to Eq. \ref{eq:wdf0}. The different iso-$k_x$ lines correspond to the diffraction orders (red lines) and intermodulation terms (blue lines). 
Passage of the wave through the OTE is conventionally modeled by the multiplication of the wave function and the OTE transfer function in real space. In phase space, this is equivalent to multiplication in $x$ and convolution along $k_x$. Since the WDF of the plane wave is a Dirac delta in $k_x$ (Fig. 1a), the WDF after the passage resembles the WDF of the OTE. 
Further propagation over a distance $d$ applies a shear of  $\lambda k_x d/2\pi$ along $x$ axis to the WDF. Larger spatial frequencies experience a larger amount of shift in $x$ during propagation. 

The Talbot effect can be explained easily by using Eq.\ref{eq:wdf0} \cite{testorf2010phase}. The period of the intermodulation terms in the WDF is $a$ and the shift experienced by all iso-$k_x$ lines, is a multiple of the shift of the first intermodulation line ($m=1$,$n=0$). Hence to reproduce the WDF after propagation, it is enough that this line is shifted by a multiple of the period. This occurs for propagation distances that are a multiple of $z_T=2a^2/\lambda$, which is called the Talbot distance. At multiples of the Talbot distance from the OTE, self-images of the OTE form . Depending on the Fourier coefficients $t_n$, other high contrast images may be observed at fractions of the Talbot distance, which is called the fractional Talbot effect \cite{Berry}. In the far field, the plane wave practically produces diffraction orders, which can again be explained by our model. Paraxial propagation to infinity is equivalent to application of Fourier transform to the wave. In phase space this mean an exchange of $x$ and $k_x$ axes, or a $-90^\circ$ rotation. When this is applied to Fig.\ref{fig:1} (b), all lines become vertical. The intensity is obtained by intergration along $k_x$, which eliminates the oscillating intermodulation terms and leaves only the diffraction orders. 

\begin{figure}
\begin{tabular}{cc}
\includegraphics[width=0.2\textwidth]{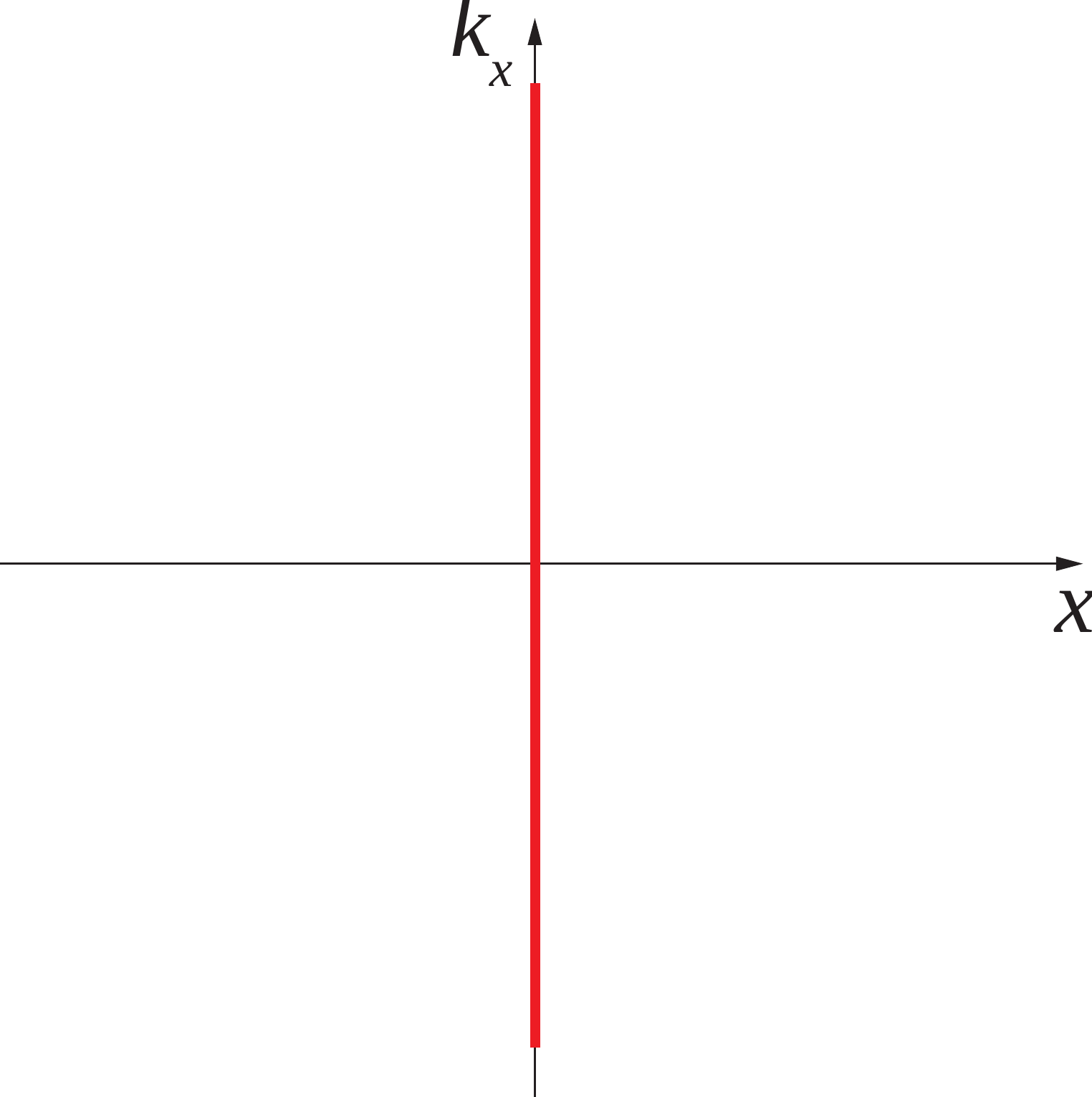} &
\includegraphics[width=0.2\textwidth]{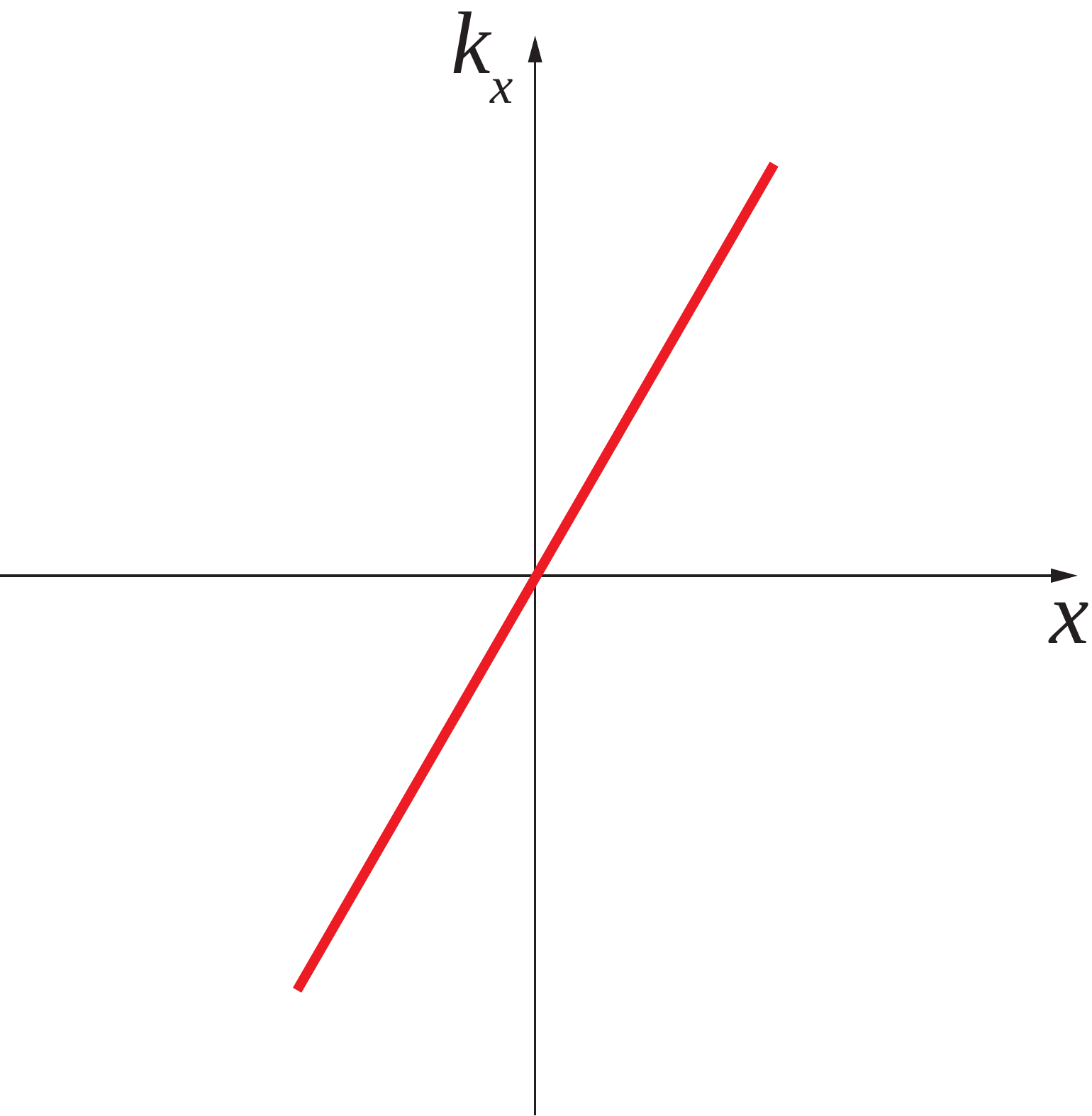}
\\
(a) & (b) \\
\includegraphics[width=0.2\textwidth]{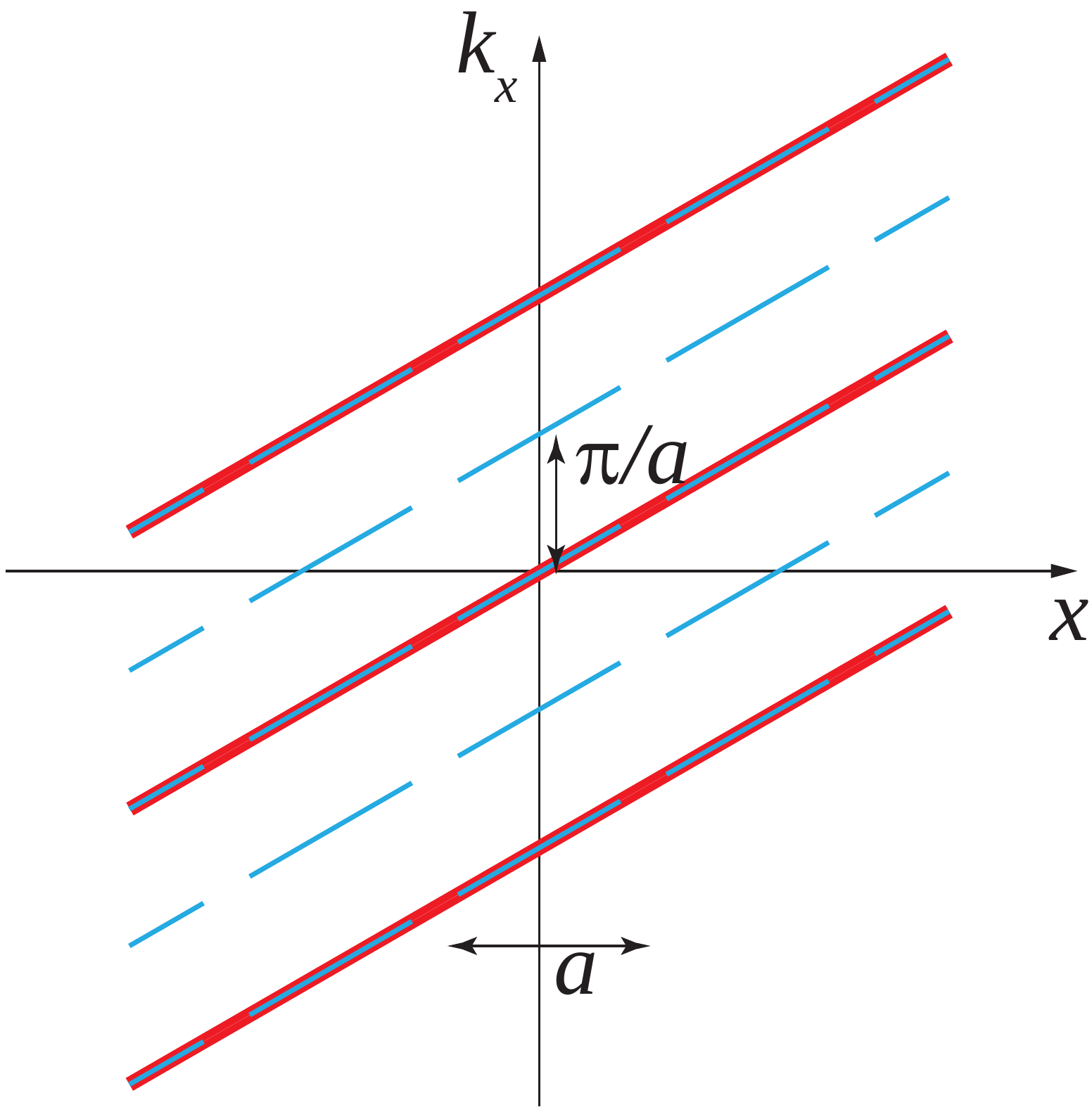} &
\includegraphics[width=0.2\textwidth]{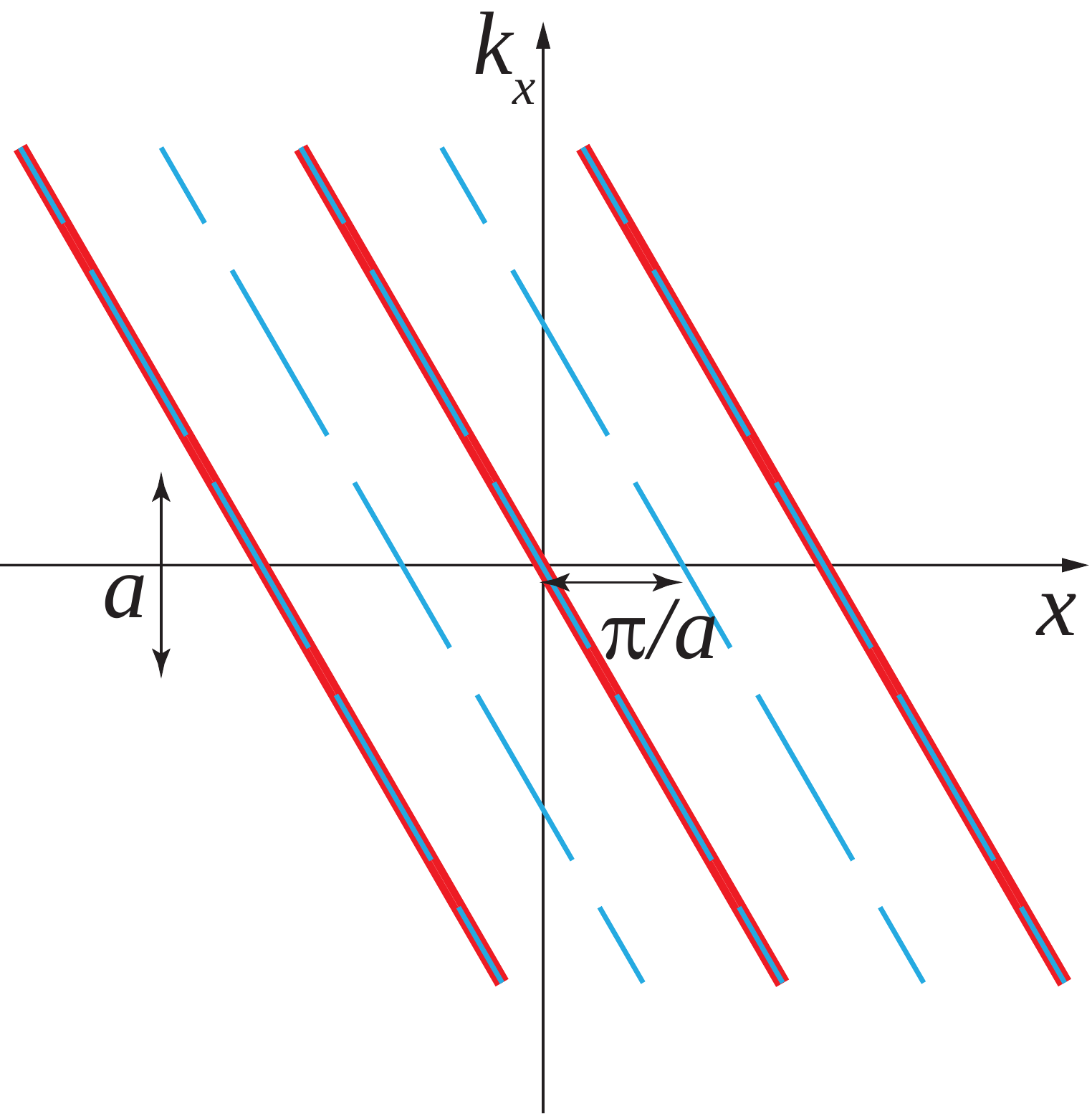}\\
(c) & (d) \\
\end{tabular}
\captionsetup{singlelinecheck=off}\caption{The WDF of (a): a point source before the primary propagation, (b): a point source after the primary propagation, (c): the point source after passage through the OTE, and (d): the image at infinity. }
\label{fig:2}
\end{figure}

Similarly it is possible to explain the angular Talbot effect \cite{testorf2010phase}, that is if the plane wave is replaced by a point source at a distance $d=mz_T$ from the OTE , high-contrast images form at infinity which are periodic in angle \cite{Azana}. We consider the point source at $z=-d$, whose WDF is a Dirac delta line $A(k_x )\delta (x)$ where $A(k_x )$ shows the wave’s spectral spread. Just before passing the OTE, the source WDF becomes an oblique line in phase space due to propagation over the distance $d$. The OTE transfer function is still explained by Eq.\ref{eq:wdf0}  in phase space. As before, passage of the light emitted by the point source through the OTE is equivalent to a multiplication with respect to $x$ and a convolution with respect to $k_x$. Convolution of the line $k_x=k_{x_0}$ and the tilted line shifts the tilted line in $k_x$ by $k_{x_0}$ which results in the WDF depicted in Fig.\ref{fig:2} (c). Only a few lines are shown to simplify visualization. The WDF at infinity is obtained by a $90^\circ$ rotation of the WDF after passing the OTE (Fig\ref{fig:2} (d)). The $x$-dependent part of the intermodulation terms in Fig\ref{fig:2} (c) are in phase, which means that they have the same shadow on the $x$ axis. To produce a self-image at infinity, these terms should also have the same shadow on the $k_x$ axis because the final image is obtained by a $90^\circ$ rotation of the WDF. This implies $\frac{d}{k_{0}}\frac{\pi}{a}=ma$, or equivalently $d=mz_T$. The image is periodic in angle due to exchanging the $x$ and $k_x$ axes in the final step (Fig\ref{fig:2} (d)). 

\begin{figure}
\begin{tabular}{cc}
\includegraphics[width=0.2\textwidth]{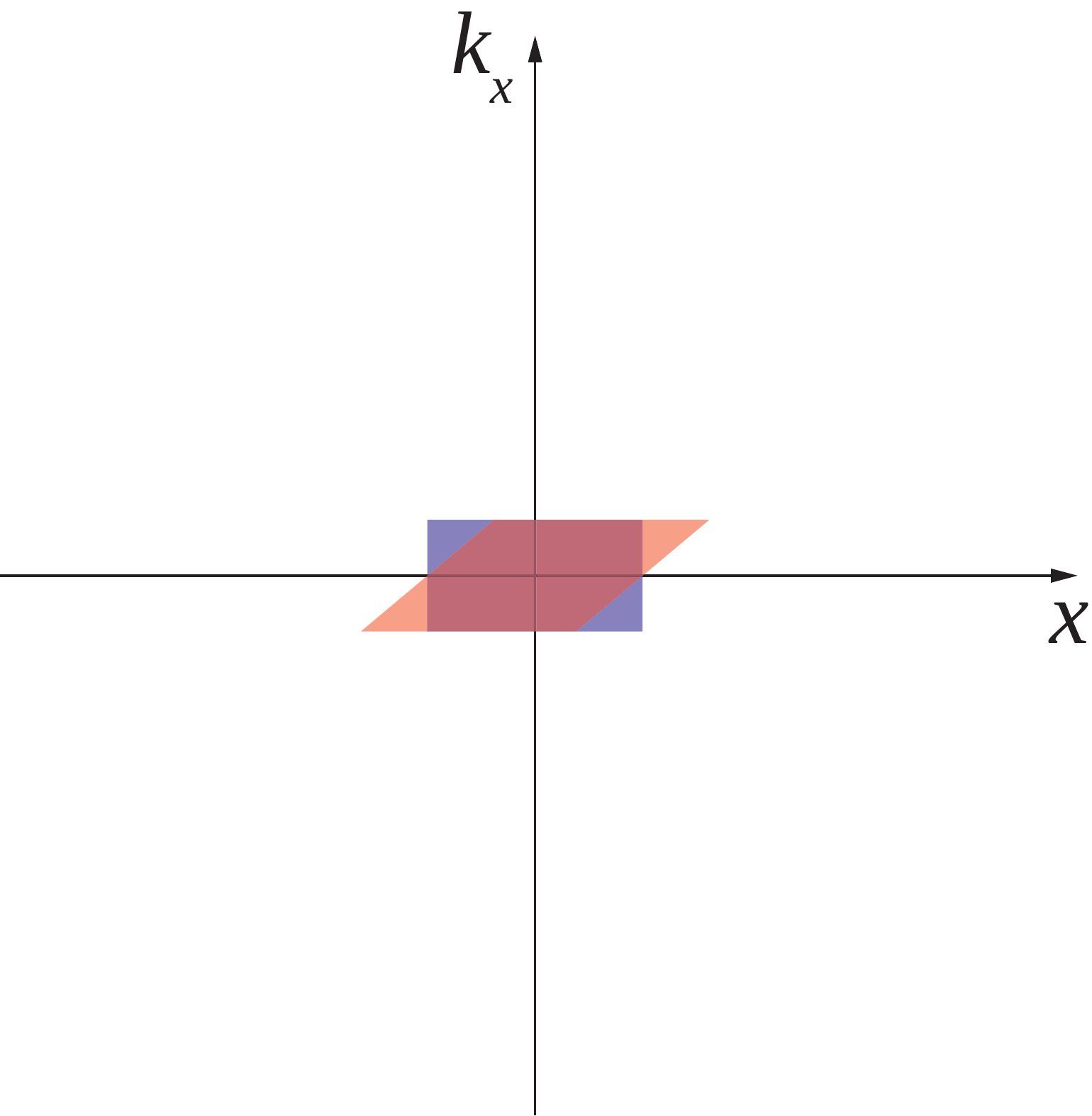} &
\includegraphics[width=0.2\textwidth]{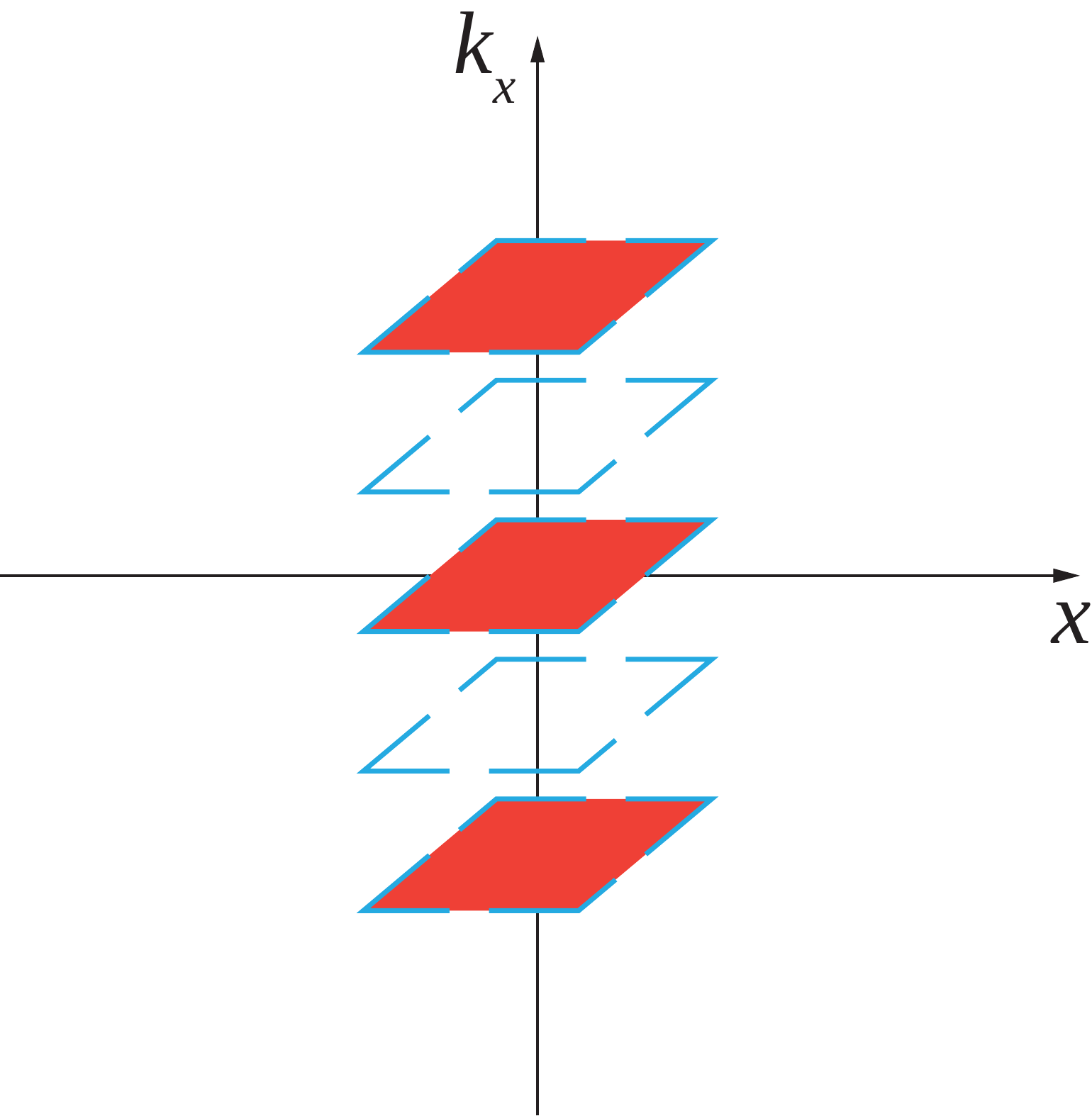}
\\
(a) & (b) \\
\includegraphics[width=0.2\textwidth]{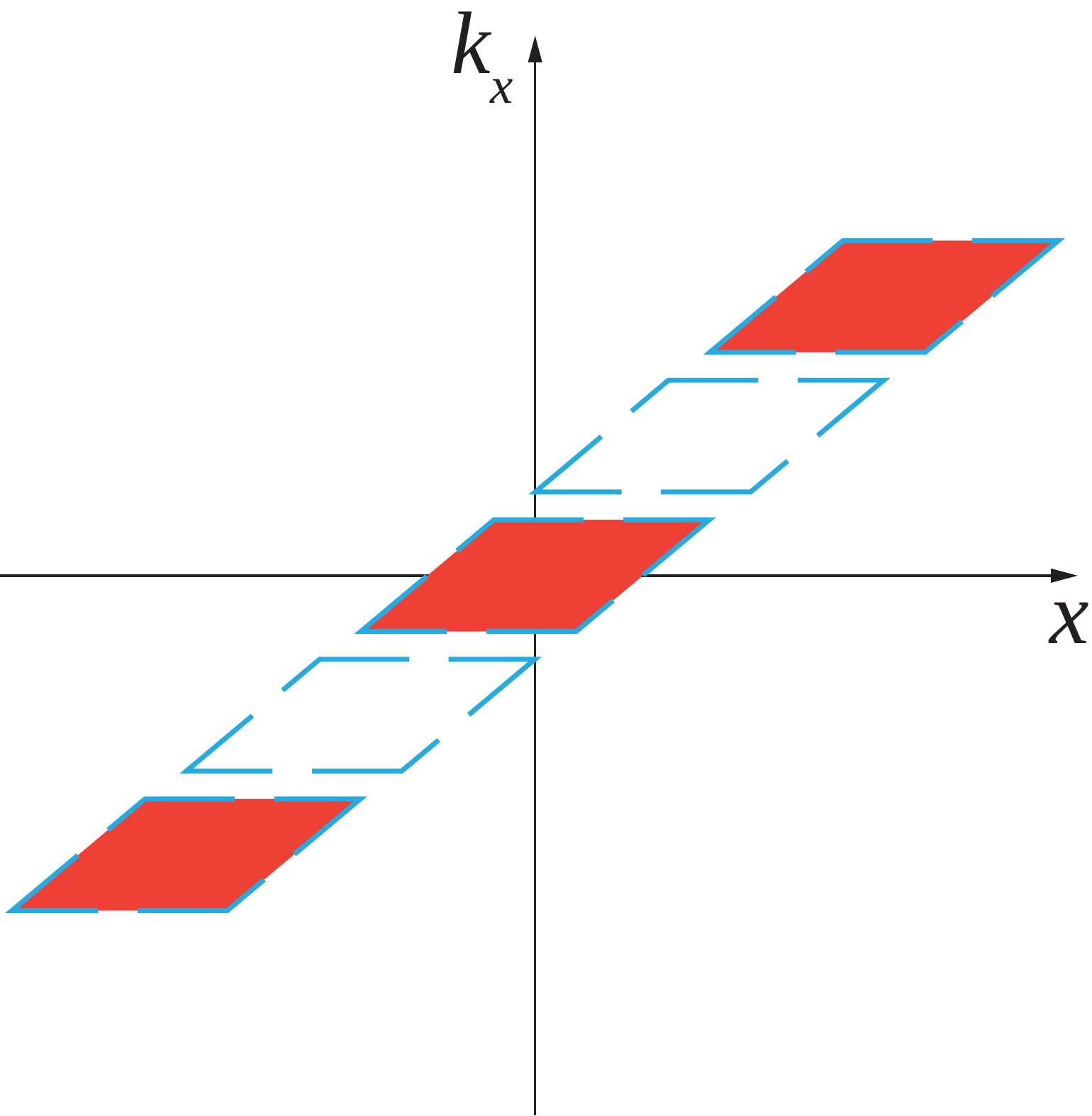} &
\includegraphics[width=0.2\textwidth]{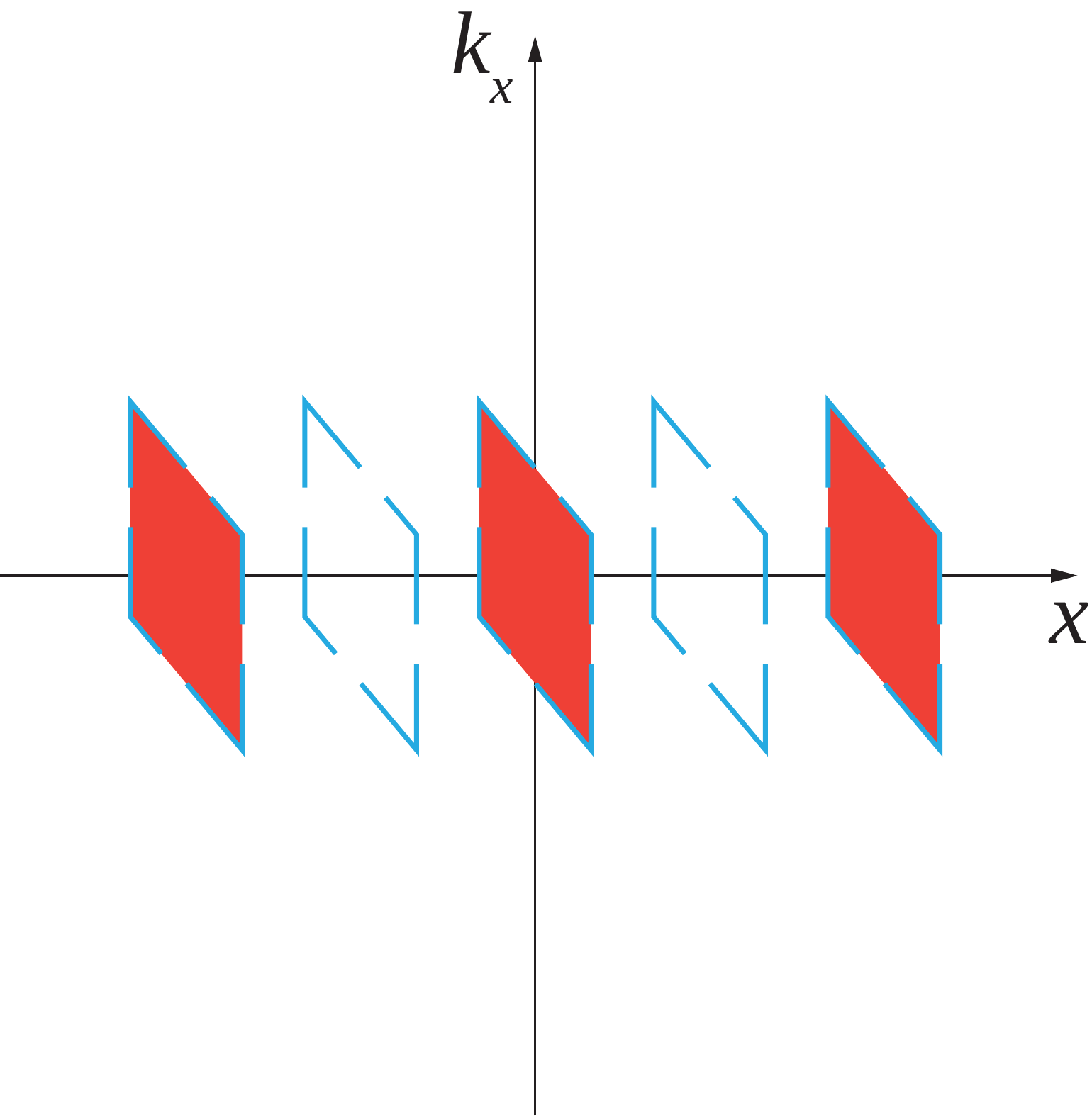}\\
(c) & (d) \\
\end{tabular}
\captionsetup{singlelinecheck=off}\caption{The WDF of (a): a source with finite spatial extent (blue) and the source after the primary propagation (red), (b): the source right after passage through the OTE, (c): the source after a secondary propagation, and (d): the image at infinity. }
\label{fig:3}
\end{figure}

We now consider a source with finite spatial extent that emits coherent light, and study it in phase space as depicted in Fig.\ref{fig:3}. 
The source is schematically represented by a rectangle with side lengths $\Delta x$ and $\Delta k_x$. The propagation over a distance $d$ transforms the source WDF into a parallelogram (Fig\ref{fig:3} (a)). The passage through the OTE results in replication of the WDF along the $k_x$ axis (Fig\ref{fig:3} (b)). The secondary propagation shears the resultant WDF again (Fig\ref{fig:3} (c)). If the image at infinity is desired, this step should be replaced by a $-90^\circ$ rotation which brings the parallelograms on the $x$ axis (Fig\ref{fig:3} (d)). Again the spacing between each two spots, i.e. the most significant peaks in the intensity profile, is inversely proportional to the period $a$  which shows that the image is periodic in angle. 
If the spots are two close, a high contrast image cannot be obtained even at infinity. This restricts the range of allowed periods and puts an upper limit on them. 

\section{The two self-imaging regimes}
\label{sec:4}
The wave structure after passage through a periodic OTE depends on different parameters including the beam divergence, the distance between the source and the OTE, the OTE period, the transmission function spectral content, propagation distance, and the wavelength. As two special cases, we investigate the image for a very large or a very small beam waist. 

The periodic OTE that we consider in our numerical simulations is a microlens array with a hemispherical concave profile, which can be expressed as $h(x)=(a/1.1)-\Re\left\{ \sqrt{(a/1.1)^{2}-x^{2}}\right\}$, where $a$ is the period and the sign $\Re$ represents the real part operator. The factor 1.1 is used to introduce a  flat space between each two concave sections. Each period consists of a flat space with the extent $a/11$ and a concave surface with the extent $a/1.1$.  
Thus changing the period scales the lens shape in all directions uniformly. To model this microlens array we apply the thin-element approximation \cite{GoodmanFO}, i.e. we assume that compared to free space, the microlens array only introduces an additional phase shift of $\Delta\phi=(n-1)h(x)k_0$ to the incident wave at $z=0$. The refractive index of the microlens array is $n=1.51$ and $k_0$ is the free space wave number.

\begin{figure}
\begin{tabular}{c}
\includegraphics[width=0.5\textwidth]{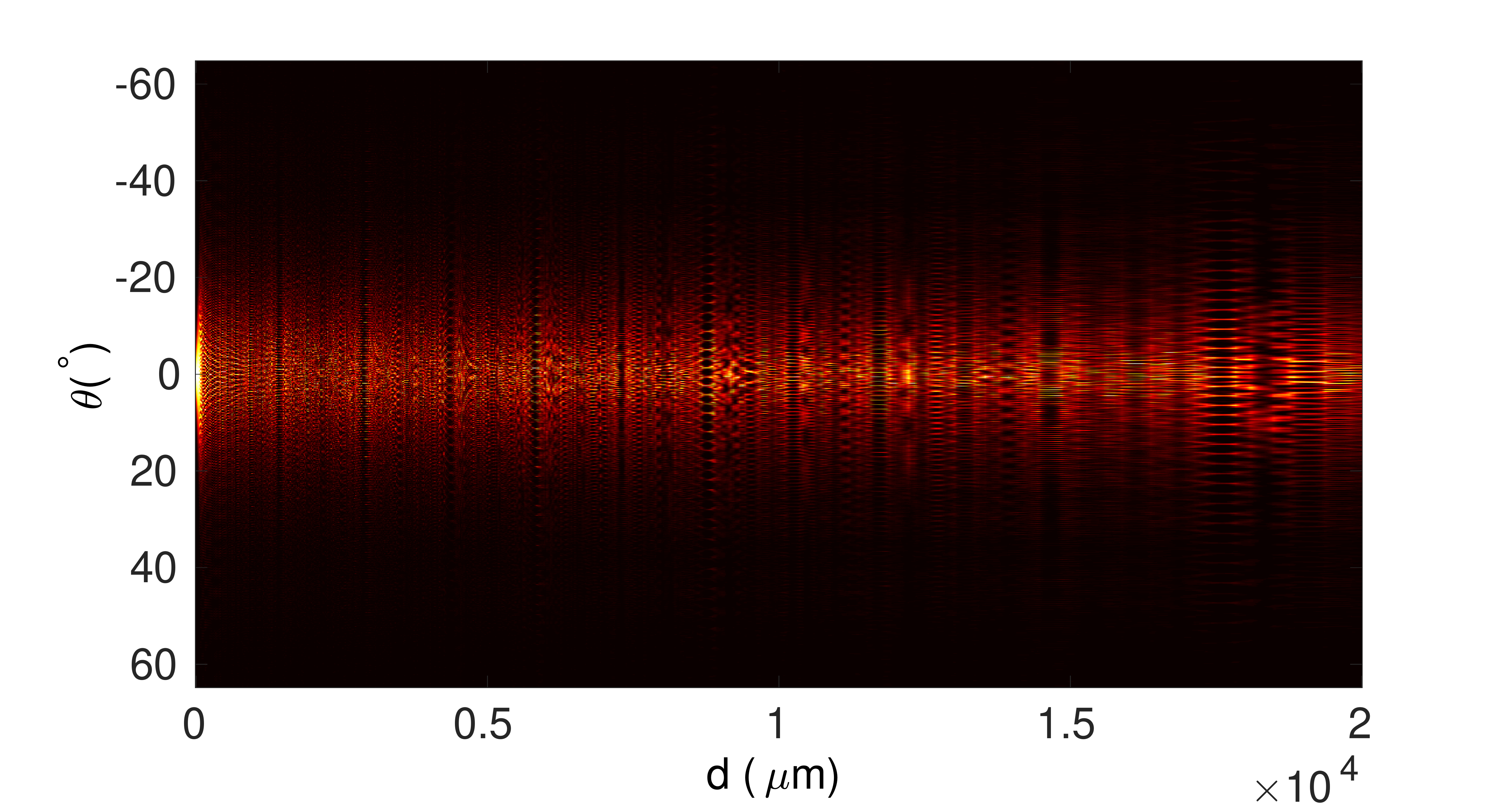}\\
(a)\\
\includegraphics[width=0.5\textwidth]{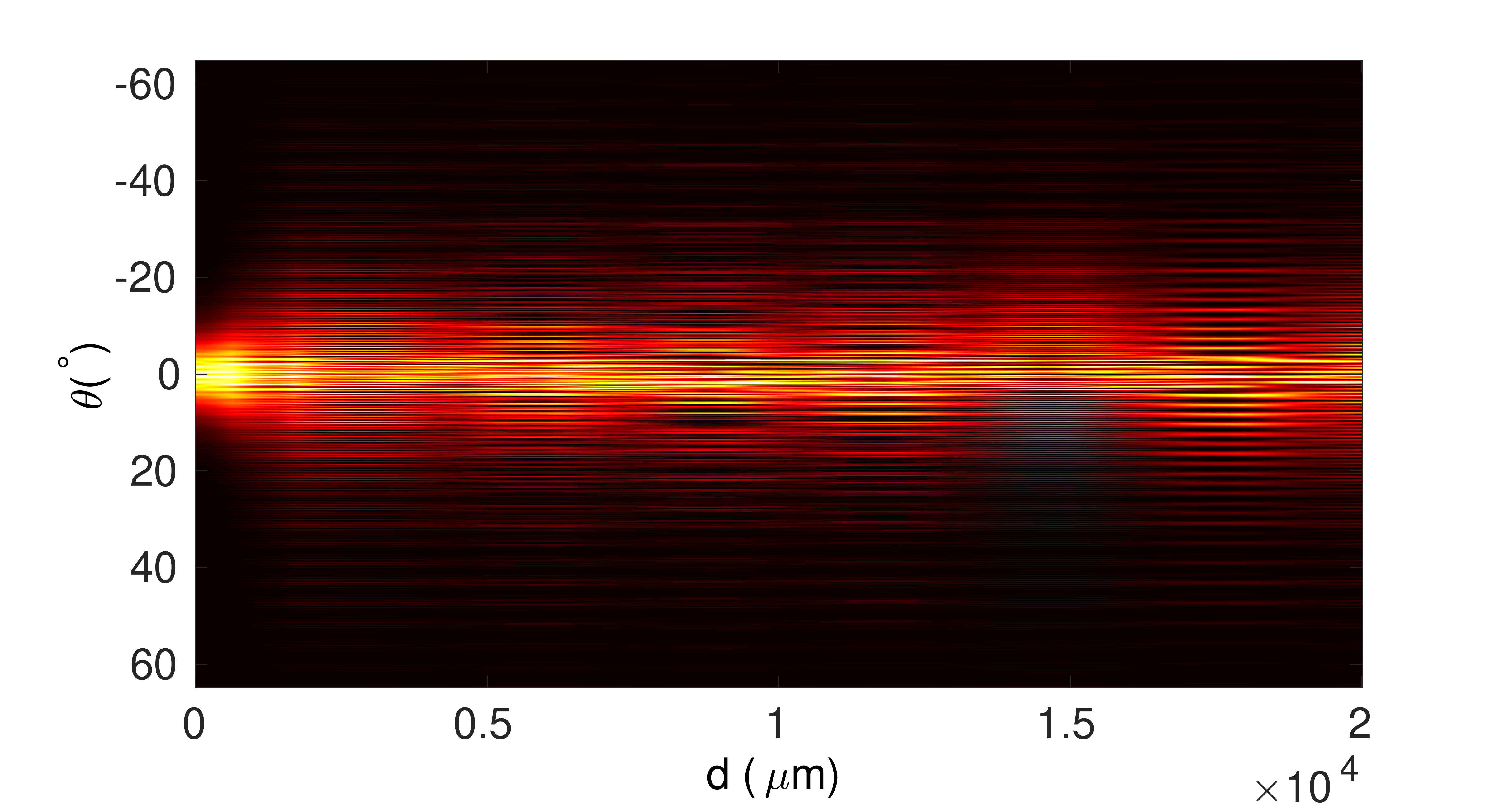}\\
(b)\\
\end{tabular}
\captionsetup{singlelinecheck=off}\caption{(a): The angular distribution of the image at infinity versus the source-OTE distance for $a=50\mu m$, $d=5.8mm$, and (a): $w_0=1\mu m$ and (b): $w_0=15\mu m$.}
\label{fig:11}
\end{figure}

We consider a source with a Gaussian beam waist of either $1$ or $15 \mu m$ at a distance $d$ from the OTE. We change $d$ to study the field profile at infinity that is obtained by applying Fraunhofer approximation. Figure \ref{fig:11} shows the intensity profiles for an OTE period of $50 \mu m$. The profile variation is much more pronounced for the smaller beam waist which provides a more divergent beam. To comapre the images obtained by the two sources, we calculate the image contrast, which is obtained by dividing the image standard deviation to its mean value, $C=\sigma/\eta$.

\begin{figure}
\includegraphics[width=0.5\textwidth]{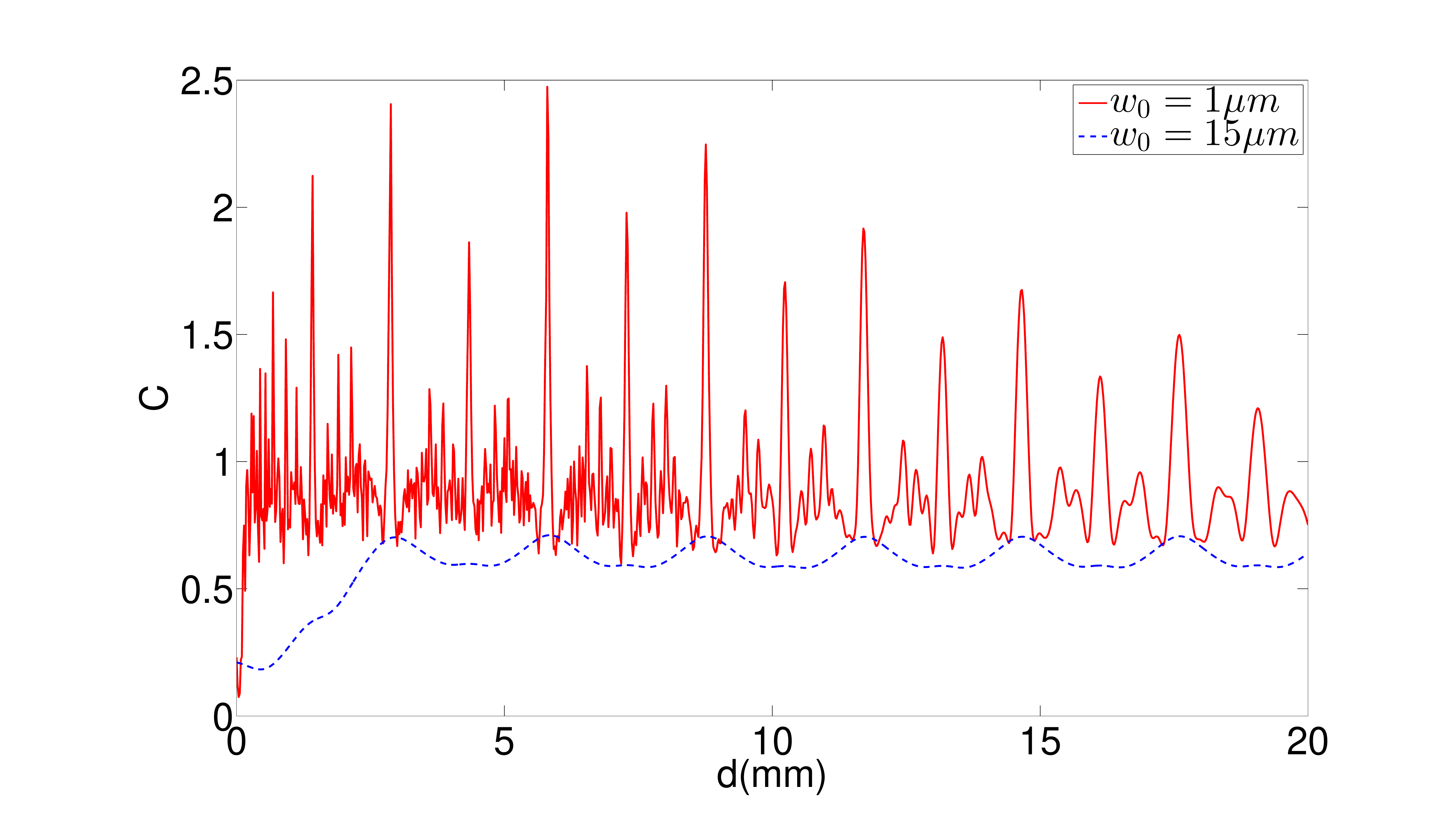}\\
\captionsetup{singlelinecheck=off}\caption{Image contrast versus the source-OTE distance for the results shown in Fig.\ref{fig:11}. }
\label{fig:12}
\end{figure}

Figure \ref{fig:12} shows that the contrast is much larger for the source with larger divergence. This seems to contradict intuition, since usually one expects that increasing the source divergence reduces contrast due to undesired interference. Another point about Fig.\ref{fig:12} is the existence of contrast peaks for the more divergent beam at distances which result in either maximal or minimal contrast for the less divergent beam. Remarkably, the second category of peaks provide lower contrast compared to the first group. The maximal contrast is obtained at $d=z_T$ for the highly divergent beam, but for the less divergent beam, contrast has the same value at all peaks. 
The results shown in Fig.\ref{fig:12} suggest a deeper analysis of the contrast. 
In the rest of this paper, we explore this point further, to provide a more complete description of light behavior after passage through an OTE.

\begin{figure}
\begin{tabular}{c}
\includegraphics[width=0.5\textwidth]{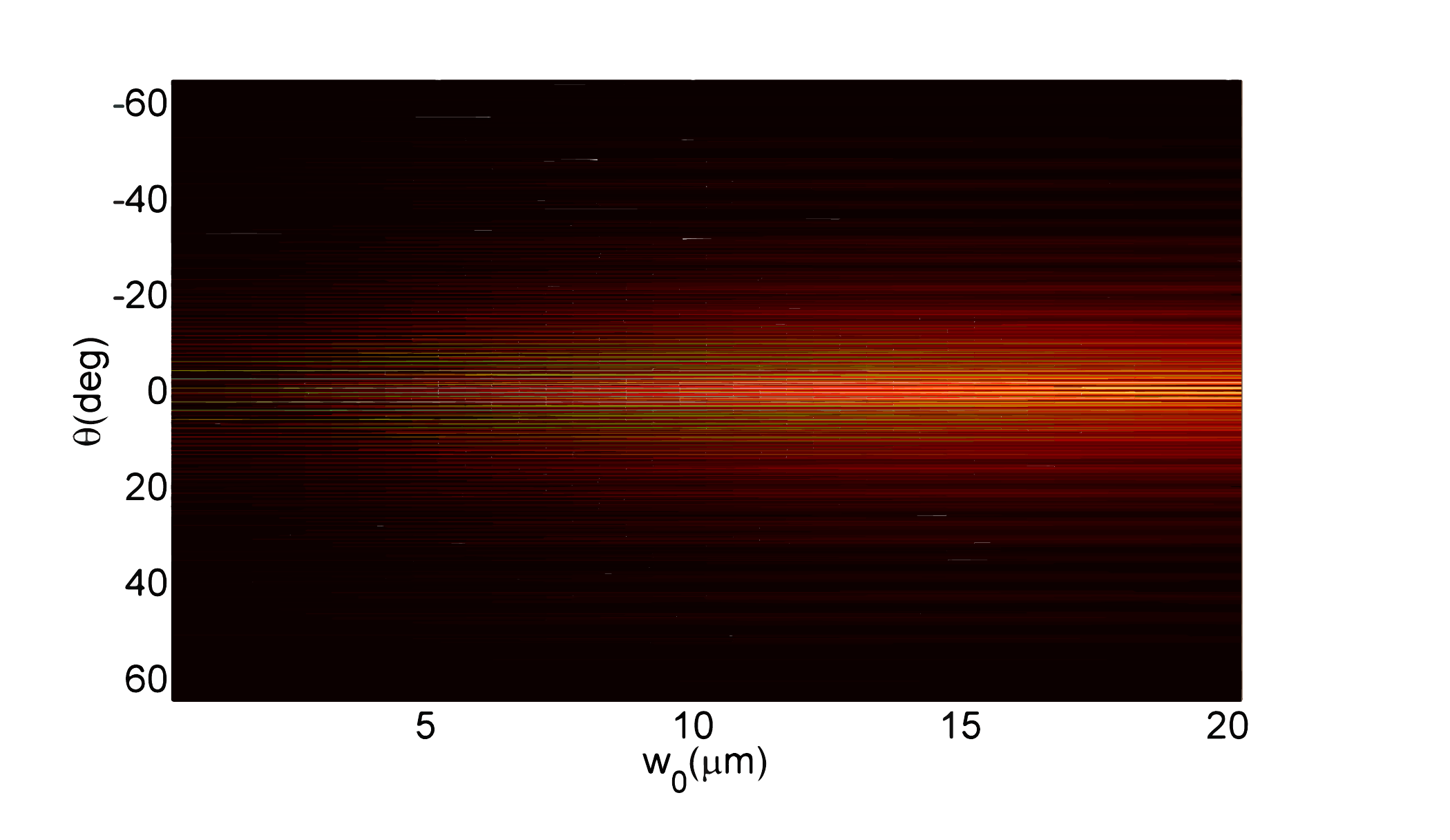}\\
(a)\\
\includegraphics[width=0.5\textwidth]{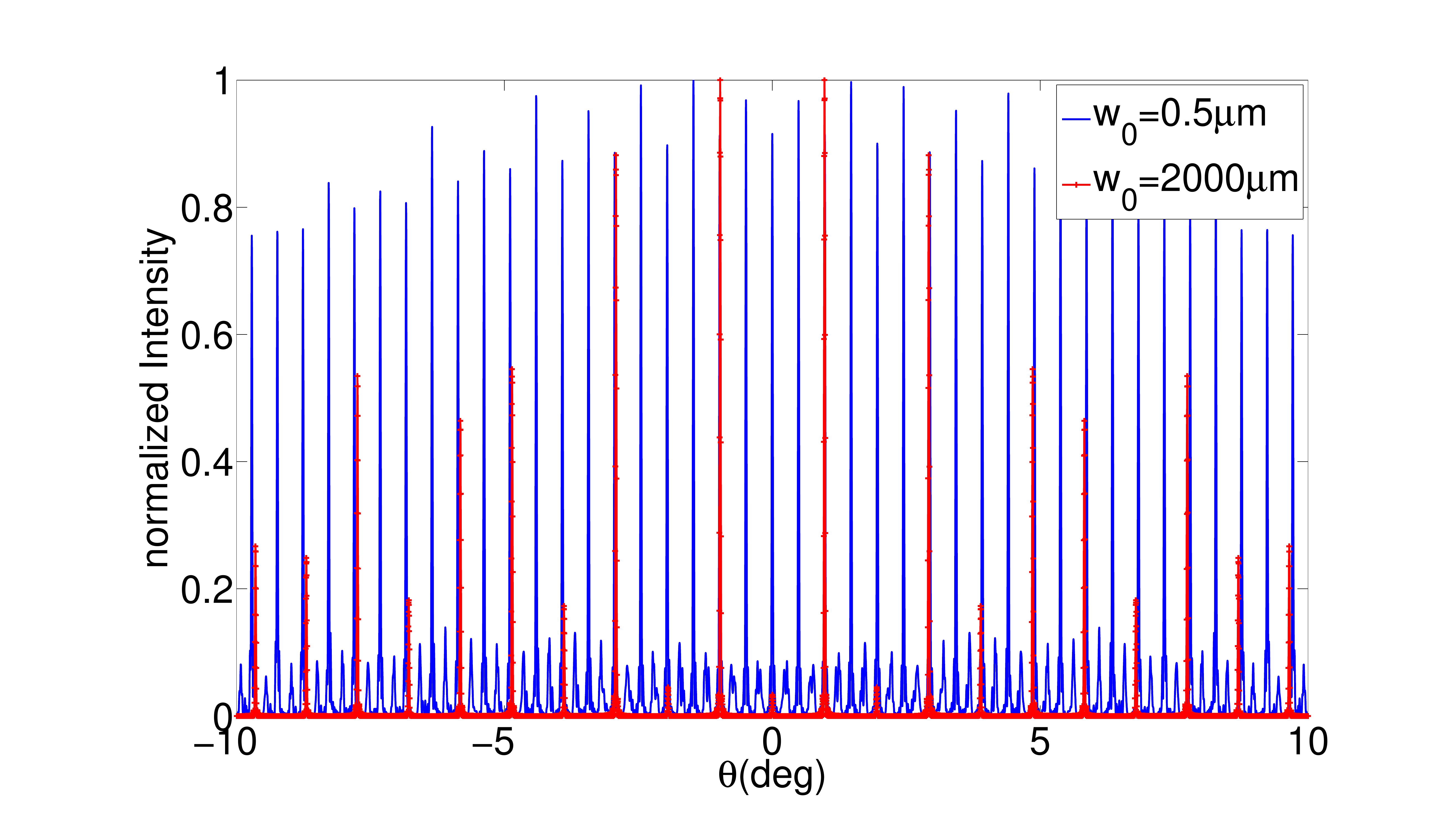}\\
(b)\\
\includegraphics[width=0.5\textwidth]{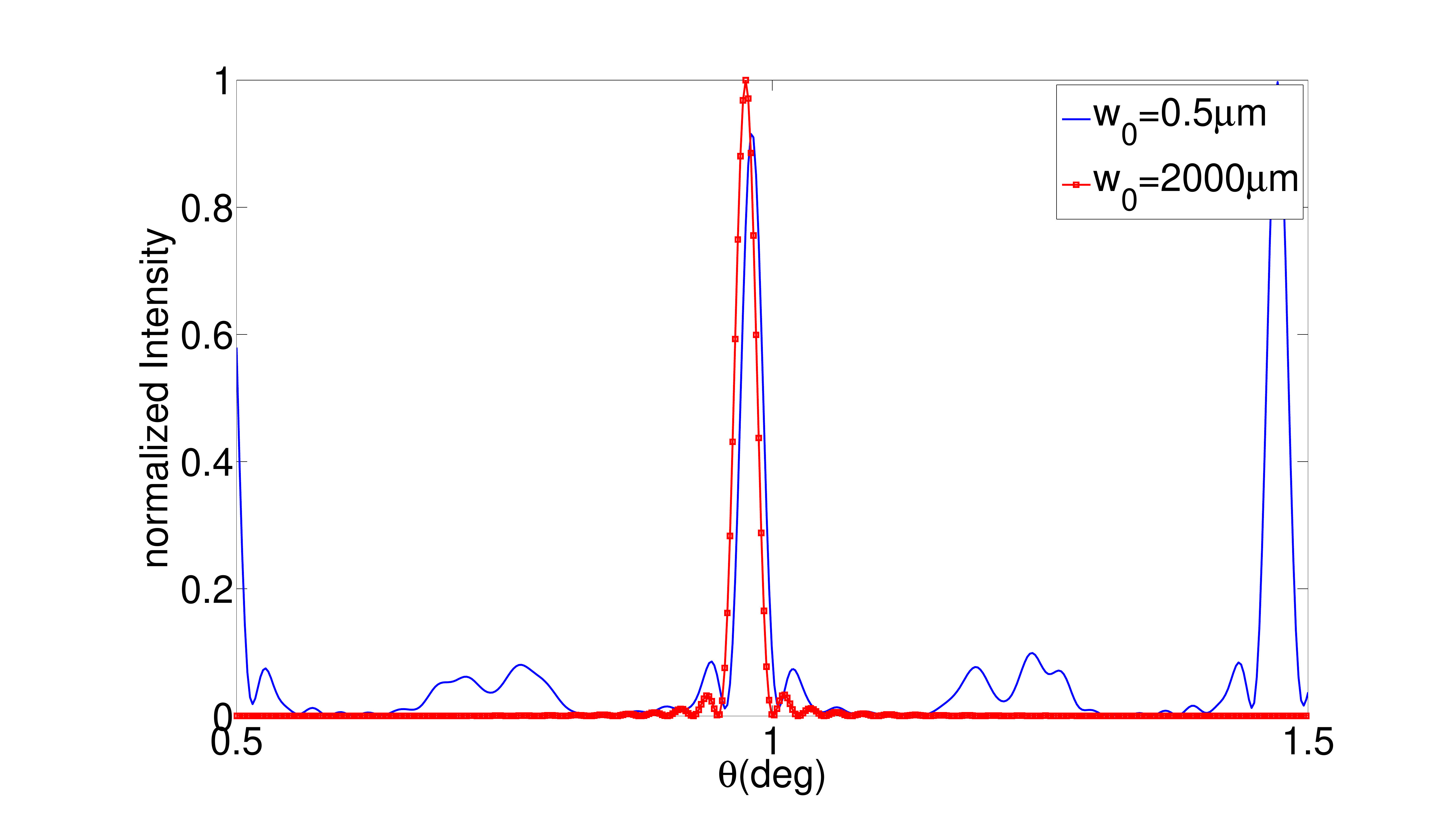}\\
(c)\\
\end{tabular}
\captionsetup{singlelinecheck=off}\caption{(a): The intensity at infinity as a function of th beam waist for parameters $a=50\mu m$, $d=5.8 mm$. (b): Same as (a), plotted for $w_0=0.5\mu m$ and $w_0=2mm$. (c): Same as (b) zoomed at around $\theta\approx1^\circ$.}
\label{fig:13}
\end{figure}

The latter example shows that the obtained contrast can depend significantly on the beam divergence. To provide a better understanding of this matter, we change the beam waist between almost zero and $80\mu m$ for $d=5.8 mm$, which brings the far-field images depicted in Fig.\ref{fig:13} (a). The source-OTE distance is chosen to be almost equal to $z_T$ to obtain high contrast due to self-imaging. 
Based on Fig.\ref{fig:13} (a) two regimes of high-contrast imaging can be distinguished; at a very small beam waist and at a very large beam waist. 
We call these two regimes the \textit{point-source} and the \textit{plane-wave} regime respectively. 
In the intermediate regime, the pattern is not as clearly pronounced as in the two extremes, thus contrast is not as high in it as in them. 
The image at infinity is different in the two extreme regimes but in each of them, it keeps its general properties and shape as the beam waist changes. The far field intensity pattern is shown in Fig.\ref{fig:13} (b), for two beam waist values of $0.5$ and $2000\mu m$, which occur at the extremes of the point-source and the plane-wave regimes. At around the propagation axis ($\theta\approx0^\circ$), where the results are most reliable, changing the beam waist alters the image periodicity by a factor of four, which is beyond the number of diffraction orders of the periodic structure. Based on the grating equation, the first order diffraction occurs at $\theta=\arcsin(0.85/50)\approx1^\circ$. This corresponds well to the image of the wide-source beam. However, for the narrow beam, the first intensity peak occurs at around $\theta\approx0.25^\circ$. Figure \ref{fig:13} (c) shows the angular extent of the spot at around $\theta\approx1^\circ$ for both cases of wide and narrow beams. Both spots have almost the same width, in agreement with the diffraction-limited nature of the system. 

\begin{figure}
\begin{tabular}{c}
\includegraphics[width=0.5\textwidth]{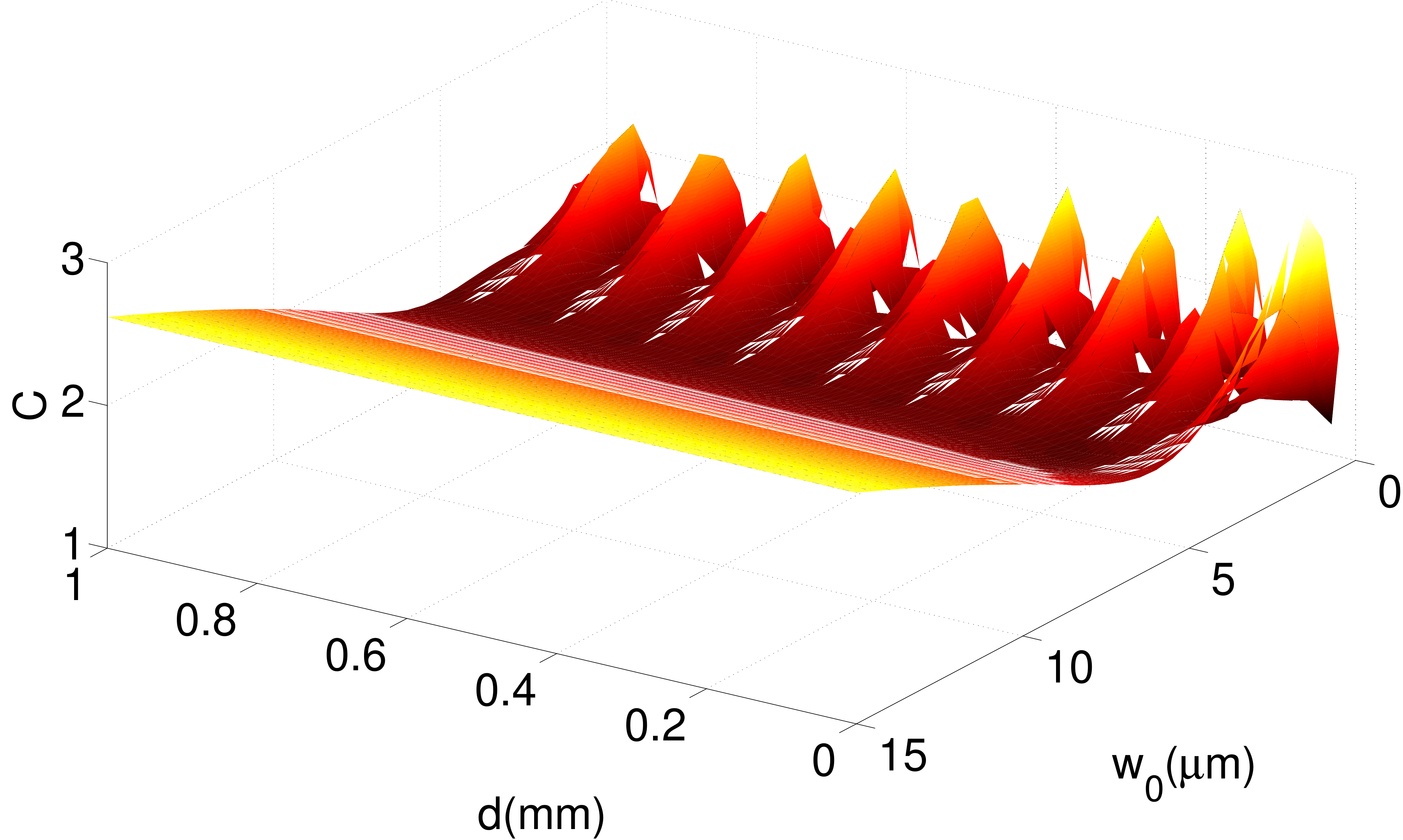}\\
(a)\\
\includegraphics[width=0.5\textwidth]{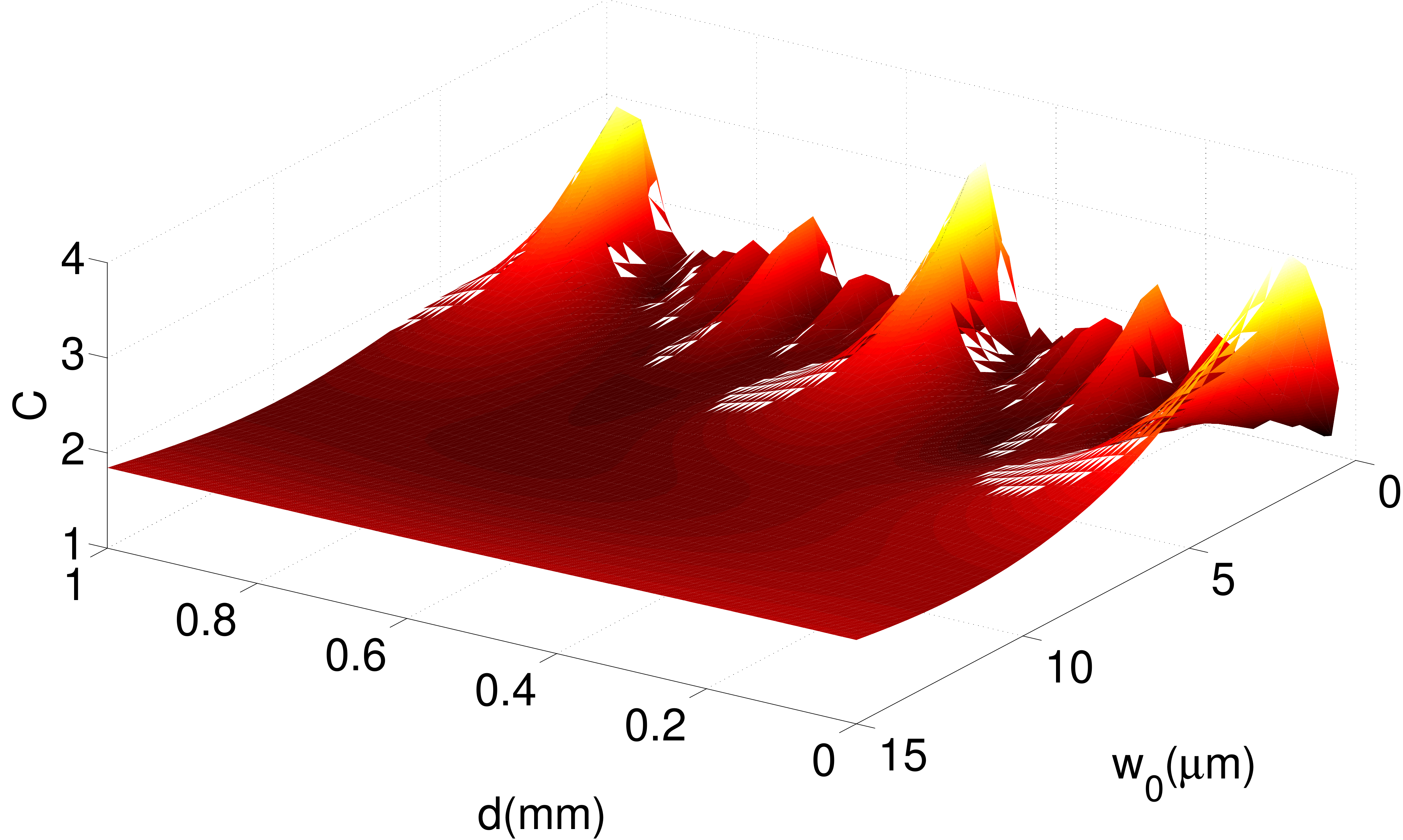}\\
(b)\\
\end{tabular}
\captionsetup{singlelinecheck=off}\caption{(a): The contrast of the image at infinity versus the beam waist and the source-OTE distance, for (a):$a=10\mu m$, and (b): $a=20\mu m$.}
\label{fig:14}
\end{figure}

Figure.\ref{fig:14} shows the calculated contrast values for the OTE periods of  $a=10\mu m$ and $a=20\mu m$, as a function of the beam waist. 
Figure \ref{fig:14} confirms the existence of the two distinct imaging regimes. These two regimes are the two regions with high contrast for small or large beam waist; in between them, the contrast is low. By changing the period from $10 \mu m$ to $20 \mu m$, the contrast peak values do not change remarkably for small beam waists. However, the contrast peaks appear less frequently and the low contrast region extends to larger beam waists in the latter case.

\begin{figure}
\begin{tabular}{c}
\includegraphics[width=0.5\textwidth]{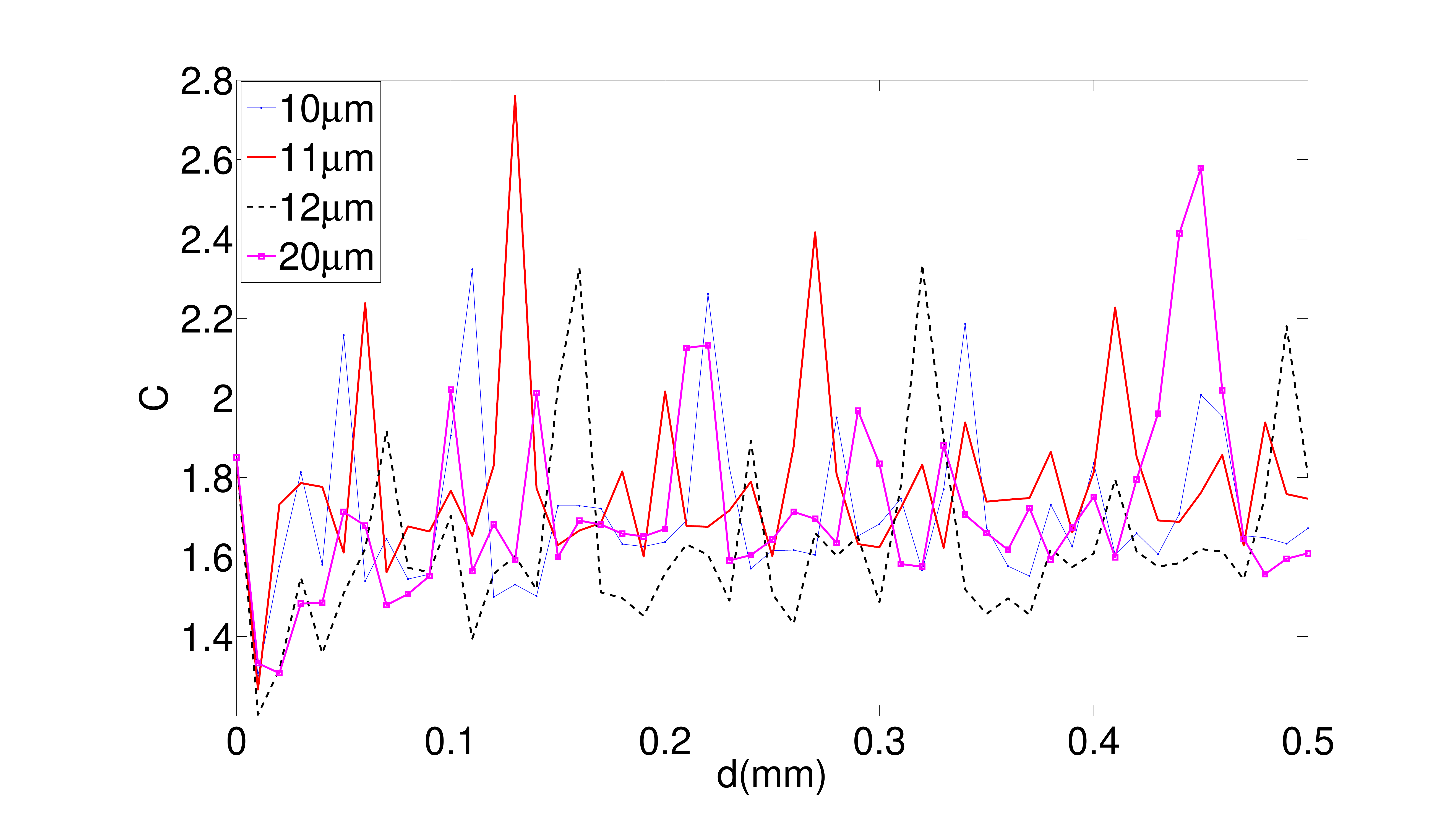}\\
(a)\\
\includegraphics[width=0.5\textwidth]{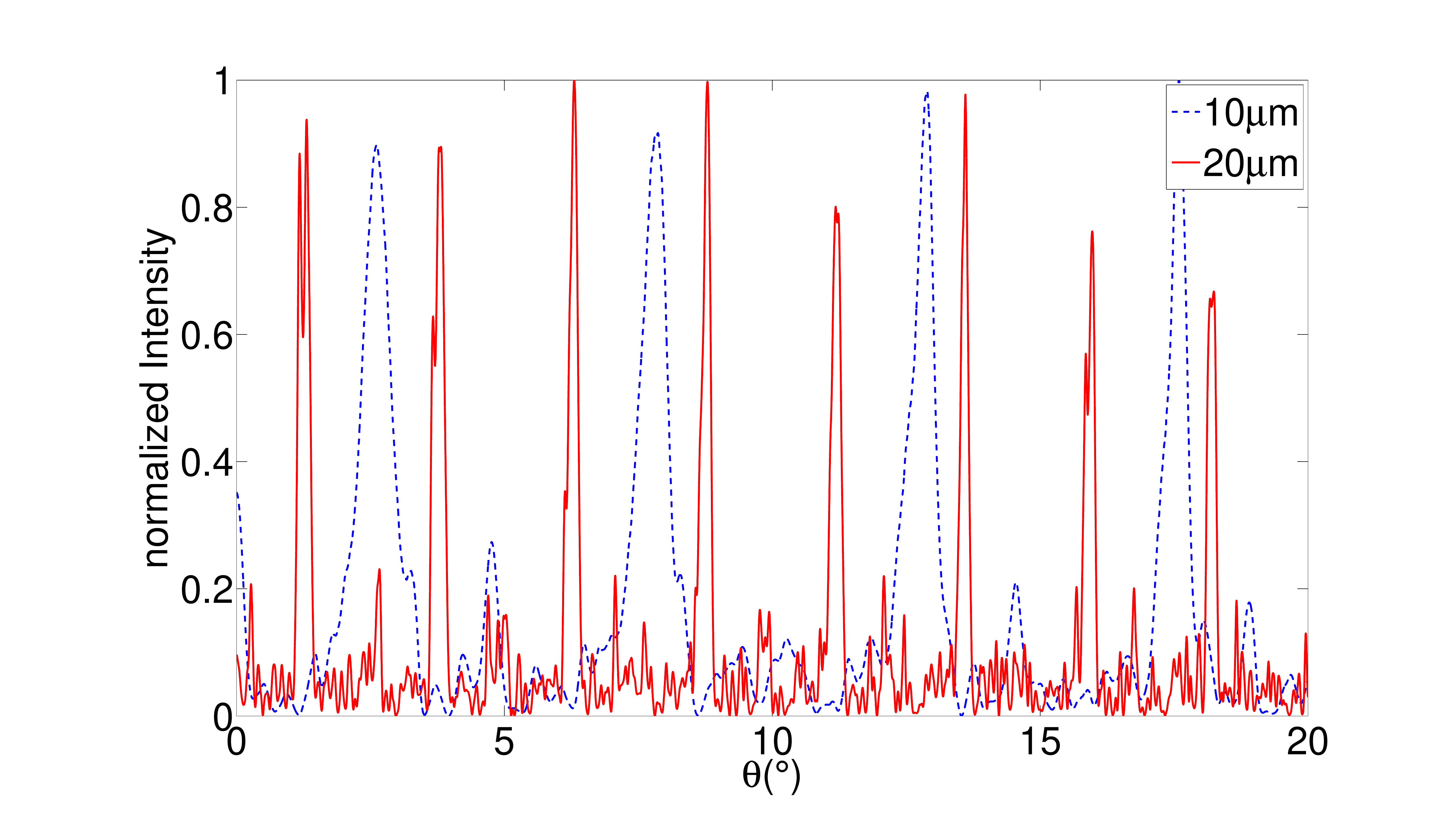}\\
(b)\\
\includegraphics[width=0.5\textwidth]{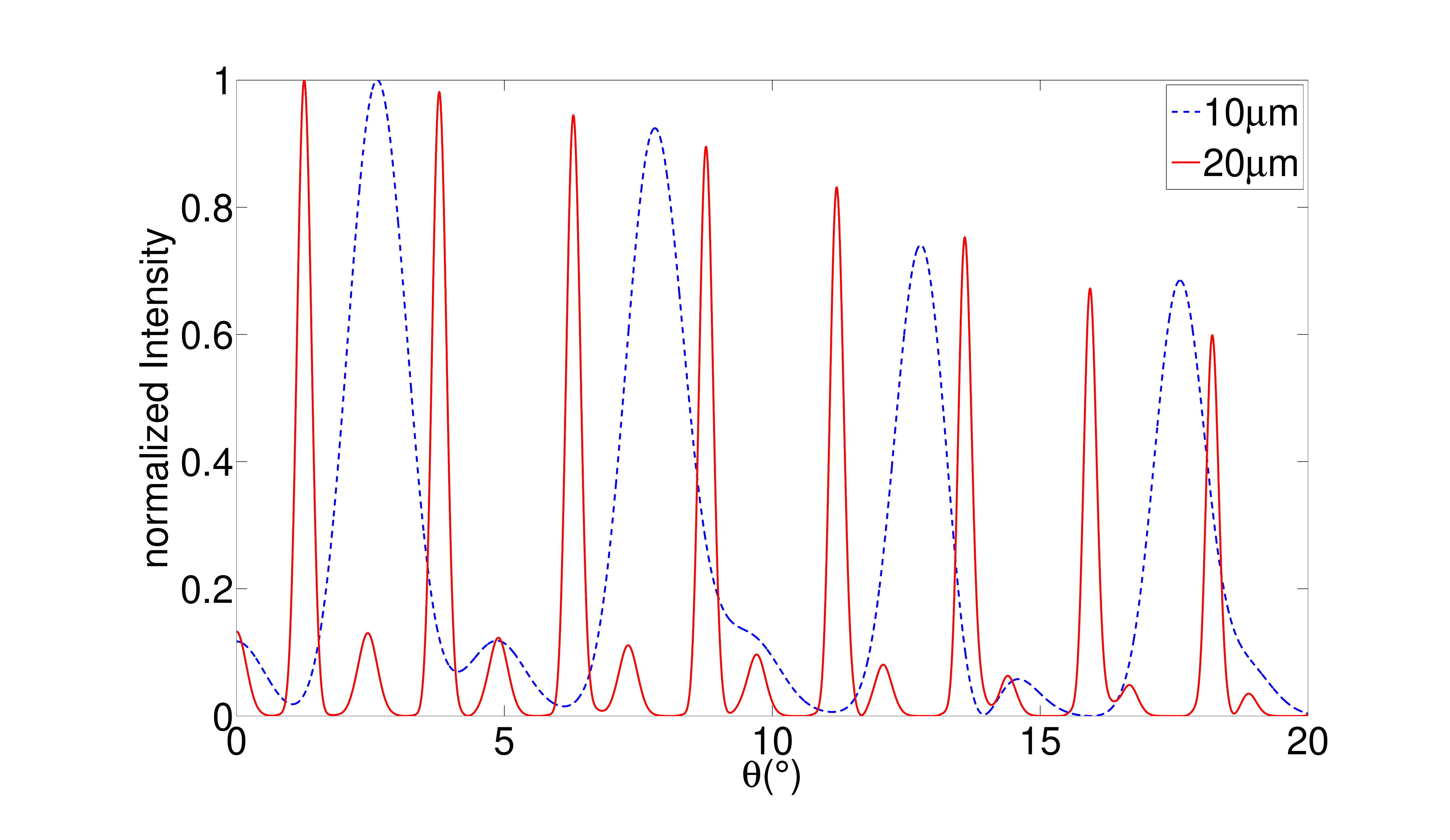}\\
(c)\\
\end{tabular}
\captionsetup{singlelinecheck=off}\caption{(a): The contrast of the image at infinity for different OTE periods versus the source-OTE distance, for $w_0=0.5\mu m$. (b): The image at infinity for $a=10\mu m$ and $a=20\mu m$. $w_0=0.5\mu m$(c): Same as (b) for $w_0=2\mu m$.}
\label{fig:15}
\end{figure}

We now aim at comparing the image in the two regimes with respect to different setup parameters. Primarily, we investigate the role of the period. Based on Fig.\ref{fig:14} we can predict that for a point-source, increasing the period leads to the less frequent occurrence of high contrast peaks as a function of the source-OTE distance ($d$ ). This is also confirmed by the Talbot distance quadratic dependence on the OTE period. To verify this more strictly, we vary the source-OTE distance for the beam waist of $0.5 \mu m$ for the periods of $10$, $11$, $12$ and $20 \mu m$. The resultant contrast is shown in Fig\ref{fig:15} (a). We take the most significant peak which occurs at $0.11 mm$ for a $a=10 \mu m$. Since $z_T\propto a^2$ , for the periods of $11$, $12$ and $20 \mu m$, the peaks should occur at $0.133$, $0.158$ and $0.44 mm$, which corresponds very well to the contrast peak locations attained in Fig\ref{fig:15} (a). We compare the image at infinity for the two periods of $10$ and $20 \mu m$ at the source-OTE distance of $0.11$ and $0.44 mm$ respectively, in Fig\ref{fig:15} (b). The number of spots for the $20 \mu m$ OTE is twice as for the $10 \mu m$ OTE in agreement with grating equation. Similar results are observed when the beam waist is increased to $2 \mu m$ as shown in Fig\ref{fig:15} (c). For the narrow beam, there is an amount of unwanted interference which leads to the appearance of noise in the image. The amount of this noise is subject to the deviation of the source-OTE distance from its optimal value. For a very small beam waist, the image becomes too sensitive to the source-OTE distance and a corresponding noise seems hard to avoid in practice, which consequently limits the achievable contrast. The width of each spot varies by changing the source beam waist. Figure\ref{fig:15} (c) shows that the beam waist is proportional the spot size, which seems nontrivial for now. We will come back to this point later in section \ref{sec:6} to explain it in phase space.

\begin{figure}
\begin{tabular}{c}
\includegraphics[width=0.5\textwidth]{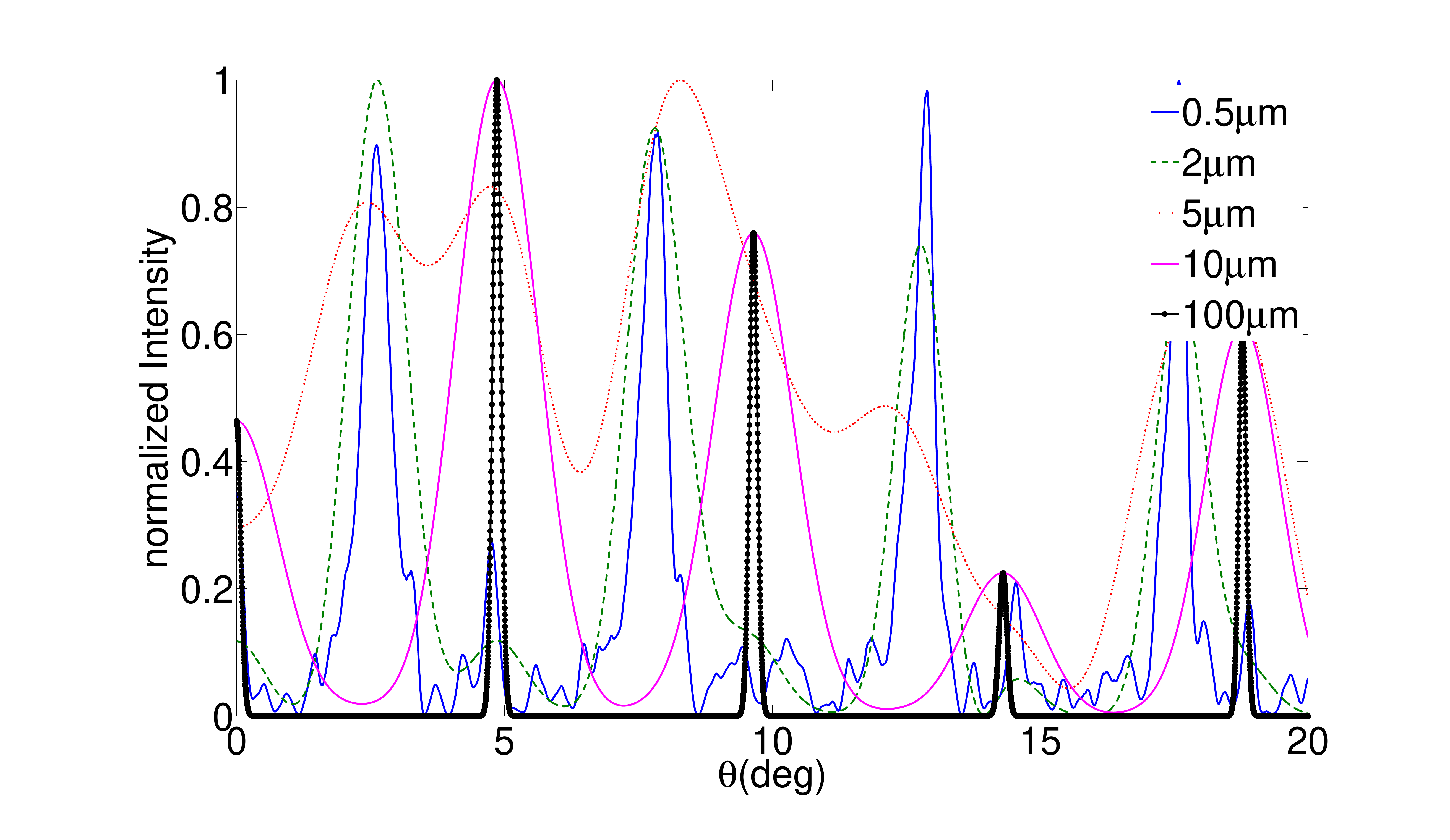}\\
(a)\\
\includegraphics[width=0.5\textwidth]{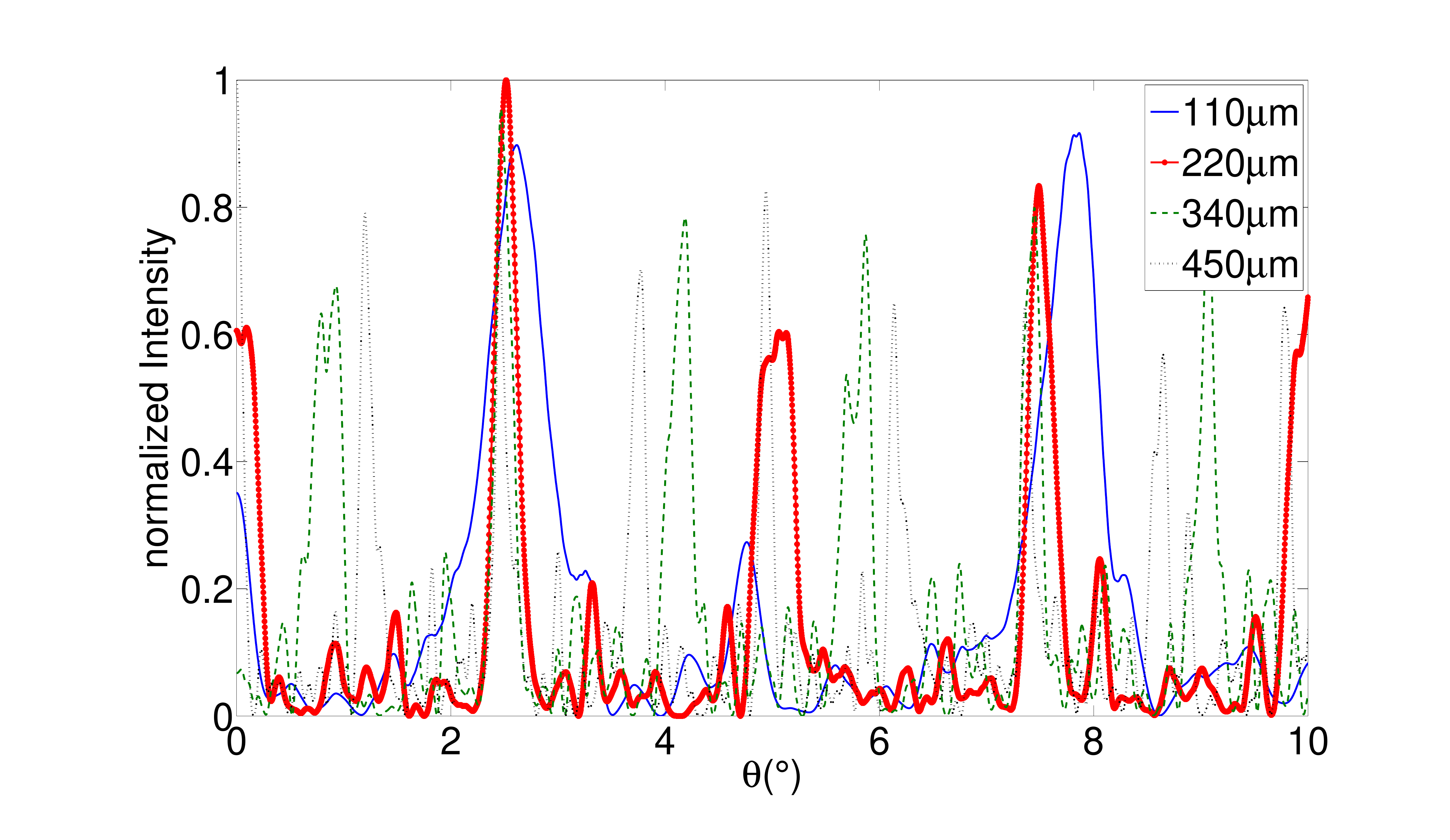}\\
(b)\\
\includegraphics[width=0.5\textwidth]{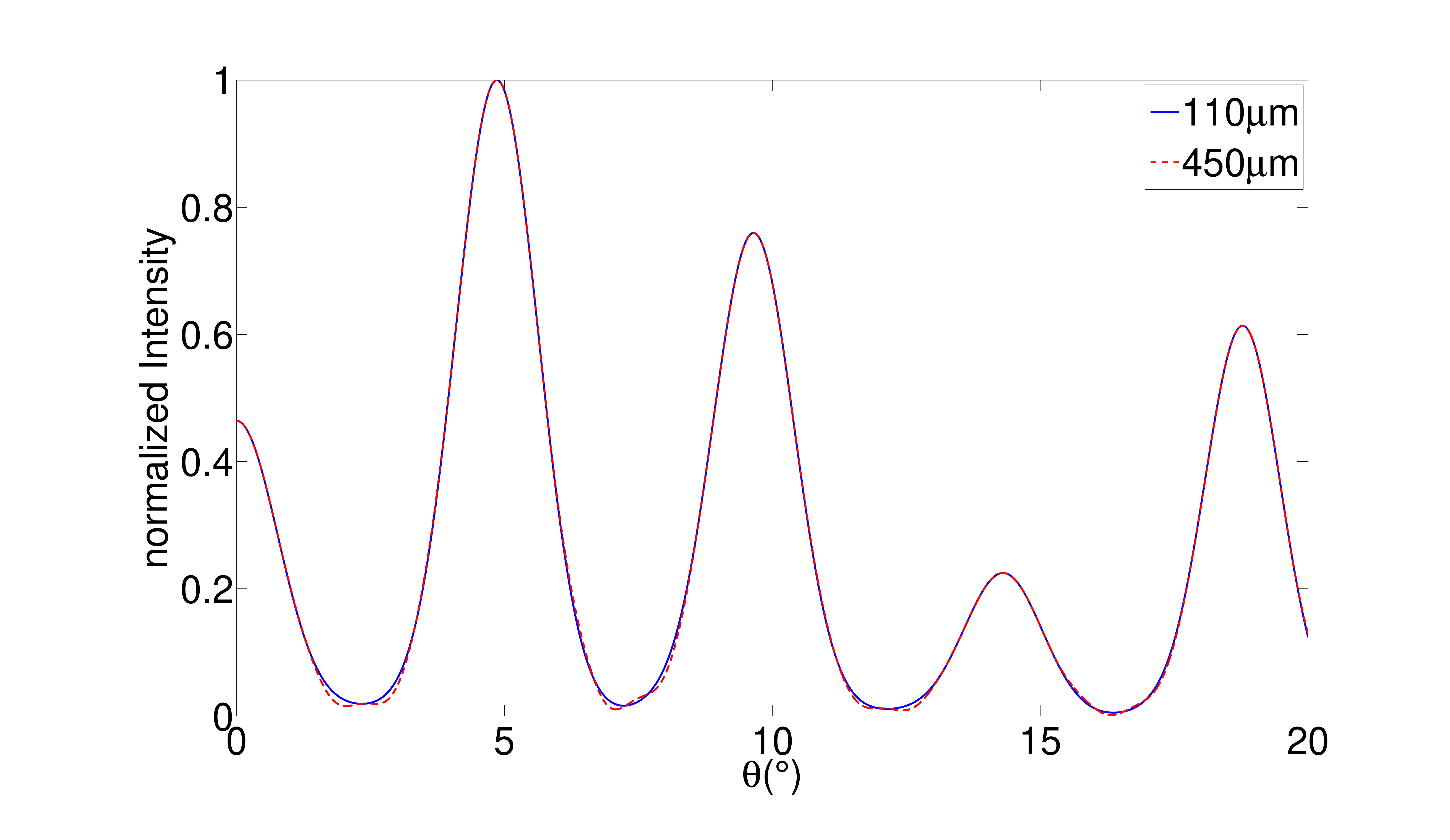}\\
(c)\\
\end{tabular}
\captionsetup{singlelinecheck=off}\caption{The image at infinity created by an OTE with a $10\mu m$ period, for (a): different source beam waists. $d=110\mu m$. (b): different source-OTE distances for $w_0=0.5\mu m$. (c): Same as (b) for $w_0=10\mu m$.}
\label{fig:16}
\end{figure}

To study the effect of beam waist on the image properties in more detail, we change the beam waist for the $10\mu m$ OTE at $d=0.11mm$. Figure \ref{fig:16} (a) shows that in the point-source regime, which occurs for the beam waist of $0.5$ and $2 \mu m$, increasing beam waist increases the spot size but in the plane-wave regime, which occurs for a beam waist of $10$ and $100 \mu m$, this trend is inversed. Locations of the peaks in the plane-wave regime is in accordance with the grating equation, contrary to the point-source regime. To connect these two regimes, there exist an intermediate regime where the peaks corresponding to both regimes exist but the contrast is reduced. The spot size is largest in the intermediate regime. For an ideal plane wave or an ideal point source, the spot size is minimal. For a point source, this is achieved at the expense of sensitivity to the source-OTE distance.  

To complete this part, in Fig.\ref{fig:16} we compare the image at the different source-OTE distances which correspond to the contrast peaks at the two beam waists of $0.5 \mu m$, which occurs in the point-source regime and $10 \mu m$, which is close to plane-wave regime. In the point-source regime, the number of spots and their width vary in direct and inverse proportion to $d$ . In the plane-wave regime changing the source-OTE distance does not alter the image. This reminds us of the observation of speckles created by illuminarting a rough surface by a laser, which may seem \textit{shifting} or \textit{boiling}, depending on whether the optical system is in the Fourier imaging regime or not \cite{GoodmanSpeckle}. 
In the plane-wave regime, shifting the object only applies a phase factor to the optical field, which is not visible as a shift in intensity. 
Contrarily, in the point-source regime, the shift of object shifts the intermodulation terms in phase space (c.f. Fig. \ref{fig:2}), which in turn leads to a shift of the image. 
This is another result of the fundamental difference between the nature of imaging for the plane-wave and the point-source regimes. 

\section{Theory for point-source imaging}
\label{sec:5}
Based on our simulations in the course of this paper, here we theoretically explain the differences between the plane-wave and the point-source illumination regimes, by using a phase space approach. Our explanation fills the gap between the predictions of the classical grating theory and our results for the point-source imaging. Specifically questions similar to the ones mentioned below should be addressed:  
\\1- How is high-contrast image formation in the point-source regime explained?  
\\2- How does theory predict the impact of different parameters in the point-source regime? Specifically why does the spot size scale proportionally to the beam waist for a point-like source?  

\begin{figure}
\begin{tabular}{cc}
\includegraphics[width=0.2\textwidth]{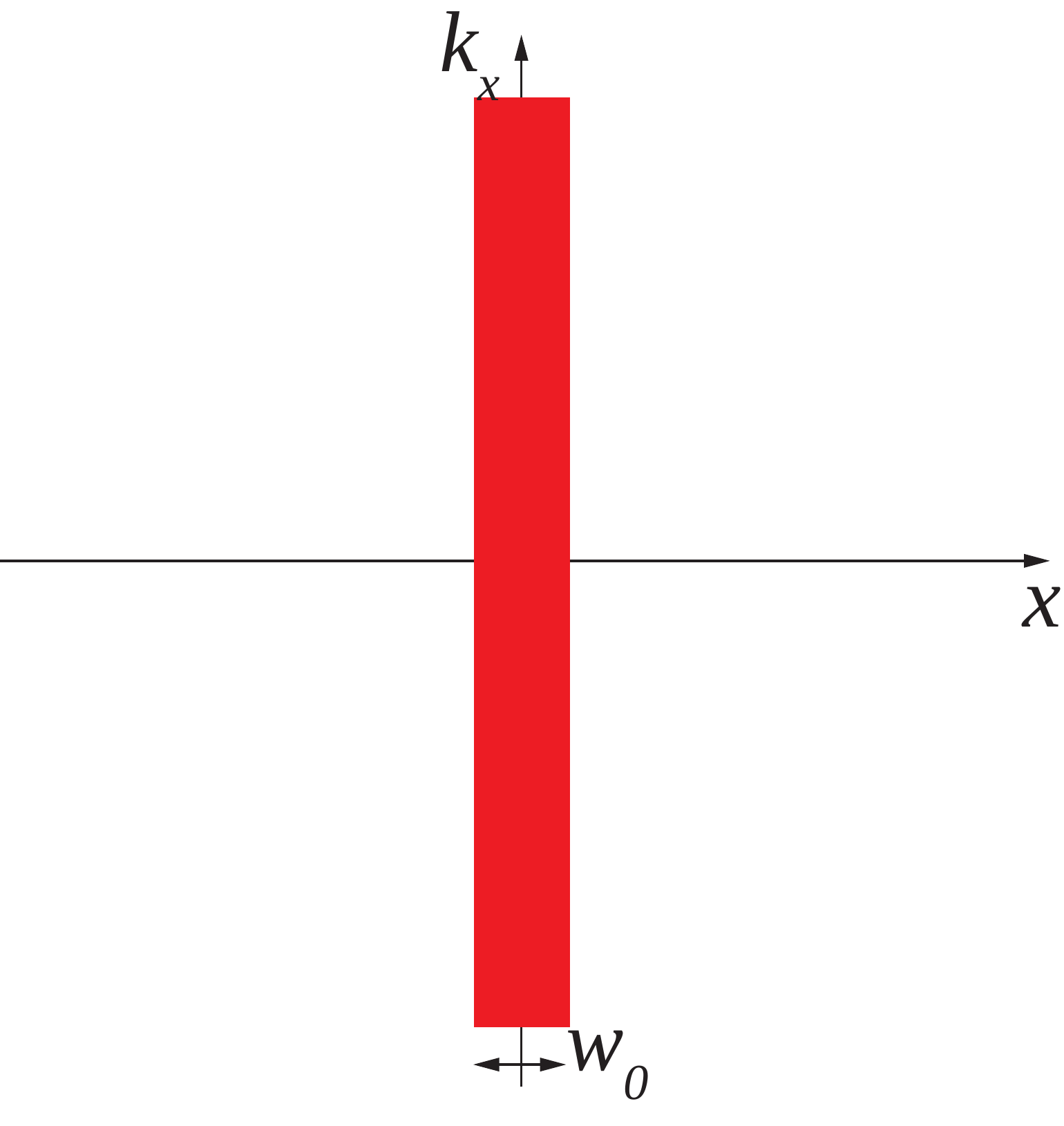}&
\includegraphics[width=0.2\textwidth]{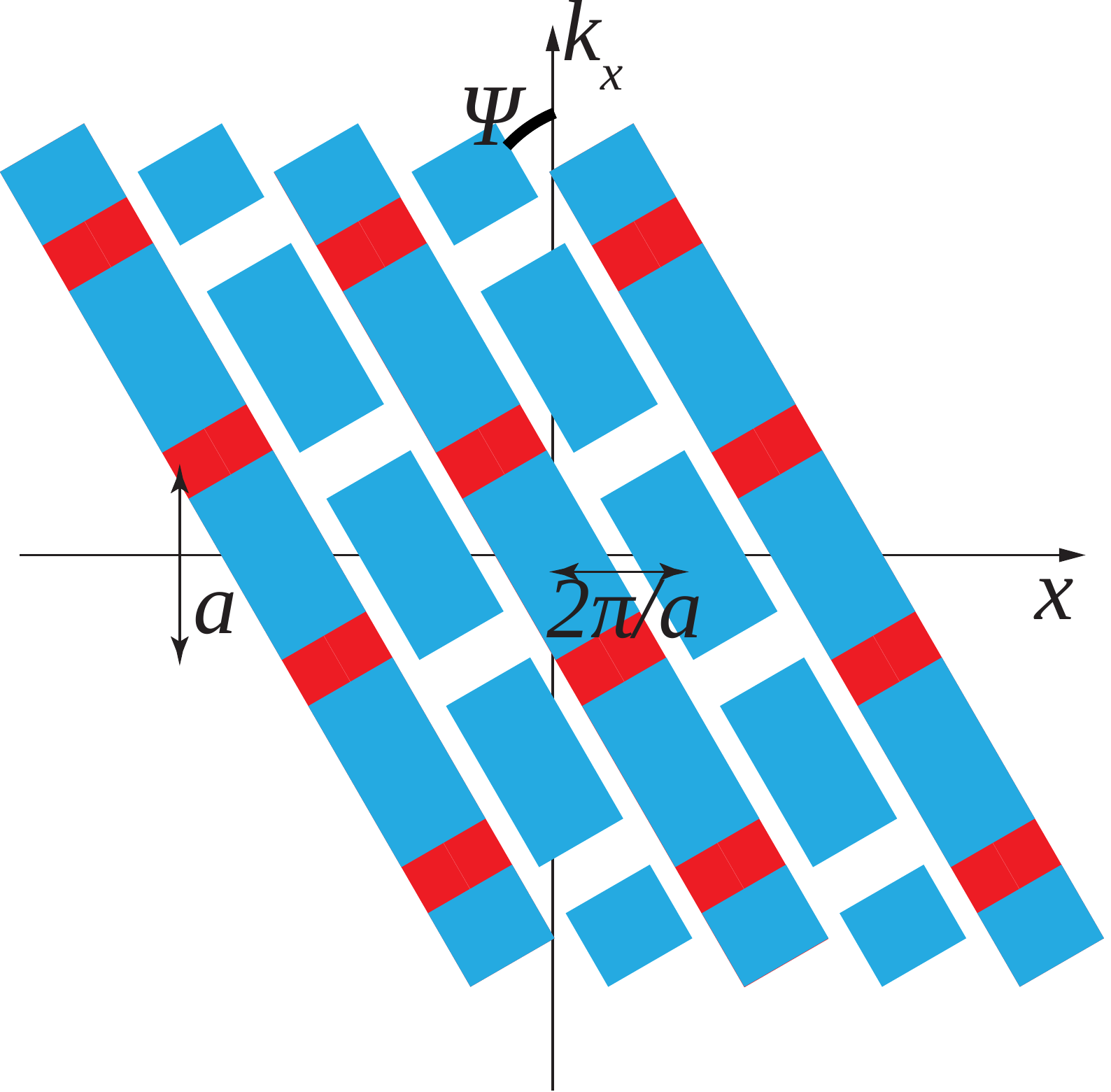}\\
(a)&(b)\\
\end{tabular}
\captionsetup{singlelinecheck=off}\caption{(a): A point source with width $w_0$ in phase space. (b): The image of the point source after passage through the periodic OTE and propagation to infinity.}
\label{fig:17}
\end{figure}

High image contrast in the point-source regime can be explained by using the phase space interpretation. A source with a beam waist of $w_0$ can be represented in phase space by a wide vertical bar as plotted in Fig.\ref{fig:17} (a). With a similar reasoning to section \ref{sec:2}, the image at infinity can be obtained, which is depicted in Fig.\ref{fig:17} (b). A smaller beam waist leads to a better separation of orders and less interference, thus provides higher contrast. In the limit of an infinitesimal point source, the source waist tends to zero and the bars become straight lines. In this way, the contradictory effect of beam waist on the contrast in the two regimes can be explained.
Similar to the case of plane-wave input, criteria can be found which should be met for high contrast imaging. To avoid interference between adjacent orders at infinity, the source waist should meet the criterion 
\begin{equation}\label{eq:pointsourceHCC}
w_{0}\sin(\Psi)<\frac{\pi}{a}.
\end{equation}
where $\Psi=\arctan(d/k_0)$. 
By using this intuition on the light behavior after passage through the OTE, the effect of different system parameters on the image can be studied. 
For example reducing the period should be in favor of high contrast because it sets the orders more apart from each other. 
In accordance with this physical intuition, a smaller period relaxes the criterion \ref{eq:pointsourceHCC} more.
Reducing the wavelength or the source-OTE distance $d$ results in smaller $\Psi$, which should in principle increase the contrast. This is in agreement with the results obtained in both Figs.\ref{fig:12} and \ref{fig:15} (a). The most significant contrast peak occurs at $d\approx z_T$. There are less significant peaks for smaller source-OTE distances but they correspond to the fractional angular Talbot effect, not the integer Talbot effect. They can thus be excluded from our discussion because in general they provide less contrast due to the inevitable existence of undesirbale intermodulation terms that do not ointerfere constructively.  
Producing more orders than the values predicted by grating equation can be understood by this phase space interpretation as well. The angle $\Psi$ is the tilt angle of the orders in Fig.\ref{fig:17} (b). For $\Psi=0^\circ$ the lines become horizontal and for $\Psi=90^\circ$ they become vertical. Increasing the source-OTE distance results in larger $\Psi$, thus leads to more vertical lines. 
The final image is determined by the summation of the all orders along $k_x$. 
To have a high contrast image, the shadow of all orders on the $x$ axis should overlap, which is the reason why the choice of $d$ is so crucial to obtain high contrast. 
The image consists of a background which consists of the shadow of both the constant (red) lines on $x$ axis and the oscillating (blue) lines. 
An additional conclusion is that due to the inevitable presence of constant orders (red lines), the ultimate contrast in the point-source regime has an upper bound that is smaller than the contrast achievable by a purely planar wave input. 
The density of spots in the image can be manipulated by changing the angle $\Psi$. For larger $\Psi$, the lines corresponding to orders become more vertical and the shadow of points become denser on $x$ axis. In this way, the system is not bound to the spot density predicted by grating equation. 
In the limit where the lines are completely vertical, which is equivalent to plane-wave input, the shadow of oscillatory lines disappears on the $x$ axis \textemdash because the integral of an oscillation over a complete period is zero\textemdash and only the constant red lines appear as spots in the image, which are separated according to grating equation. 

\section{Final remarks}
\label{sec:6}
The method and the results of the analysis presented in this paper can in principle be generalized to 2D structures. However, it will be more difficult to provide a graphical illustration because this requires a 4D space, with dimensions $x$, $y$, $k_x$ and $k_y$. Nevertheless, most of the conclusions that we derived in this paper can be generalized to the 2D lattices. 

Specifically, the WDF of a 2D transfer function can be expressed as 
\begin{eqnarray}\label{eq:wdf0_2d}
W(\mathbf{r},\mathbf{k};0)=\frac{1}{4\pi^{2}}\sum_{\mathbf{mn}}t_{\mathbf{n}}^{*}t_{\mathbf{m}}\exp\left\{ i\mathbf{r}.\left(\mathbf{g}_{\mathbf{m}}-\mathbf{g}_{\mathbf{n}}\right)\right\} \nonumber\\\times\delta\left(\mathbf{k}-\frac{\mathbf{g}_{\mathbf{m}}+\mathbf{g}_{\mathbf{n}}}{2}\right).
\end{eqnarray}
where $\textbf{r}$ and $\textbf{k}$ are vectors corresponding to in-plane position and wave-vector. Also $t_\textbf{n}$ and $t_\textbf{m}$ are the 2D Fourier coefficients of the transmission function and $g_\textbf{m}$ and $g_\textbf{n}$ are reciprocal lattice vectors. In analogy with our treatment for 1D problem, it is possible to conclude that similar effects to what we showed for 1D structures are observed for 2D lattices. In particular it is possible to derive criteria for high contrast imaging in both regimes, which we skip for brevity. 

In addition to periodic ordered elements, our method can be used to engineer light structuring by using non-periodic elements to introduce more complexity into the generated optical patterns. Periodicity necessitates the existence of nonzero WDF only for values centered around $k_x=2\pi m/a$. This condition may be in general violated by using different types of order. For example, quasi-periodic structures lead to fractal structures in $k$-space \cite{Steinhardt}. In a more general case, a random structure can be engineered by using a similar approach. In this case, the spectral content of the OTE transfer function forms a continuum and the convolution of the random OTE $k$-space representation with the wave spectral content should be shifted in phase space to model light propagation. Nevertheless, many features of light behavior after passage through the ordered structure can be modeled by using the approach that we introduced here. A special case is a phase OTE with the transfer function $t(x)=\exp (i Kx)$ with arbitrary $K$, which creates a single beam centered at $k_x=K$. 

As a final remark, we encourage the interested readers and researchers by mentioning some extensions that may complete this research. 
As we briefly discussed in the previous paragraphs, developing the method further to analyze 2D order will lead to new effects that are absent for 1D order. 
Here we studied on-axis propagation only to be able to go farther in general understanding. We believe that oblique incidence is also very interesting to investigate. Also, here we assumed that the light passes through only one optical OTE. Considering more elements should be an important problem too, and might produce new effects such as the Moir\'{e}, which are not easily achievable with one OTE, if at all possible.  
Also, a single source is considered here. It will be useful to look at a combination of sources too. 
\section{Acknowledgments}
\label{sec:7}
This work has been supported by funding in the frame of the CTI project LIMA under project number 14777.1 PFNM-NM.





 

\ifthenelse{\equal{\journalref}{aop}}{%
\section*{Author Biographies}
\begingroup
\setlength\intextsep{0pt}
\begin{minipage}[t][6.3cm][t]{1.0\textwidth} 
  \begin{wrapfigure}{L}{0.25\textwidth}
    \includegraphics[width=0.25\textwidth]{john_smith.eps}
  \end{wrapfigure}
  \noindent
  {\bfseries John Smith} received his BSc (Mathematics) in 2000 from The University of Maryland. His research interests include lasers and optics.
\end{minipage}
\begin{minipage}{1.0\textwidth}
  \begin{wrapfigure}{L}{0.25\textwidth}
    \includegraphics[width=0.25\textwidth]{alice_smith.eps}
  \end{wrapfigure}
  \noindent
  {\bfseries Alice Smith} also received her BSc (Mathematics) in 2000 from The University of Maryland. Her research interests also include lasers and optics.
\end{minipage}
\endgroup
}{}

\end{document}